\documentclass[english,final,table]{IEEEtran}
\usepackage[T1]{fontenc}
\usepackage{textcomp}
\usepackage[latin9]{inputenc}
\usepackage{xcolor}
\usepackage{amsmath}
\usepackage{amsthm}
\usepackage{amssymb}
\usepackage{stmaryrd}
\usepackage{graphicx}
\PassOptionsToPackage{normalem}{ulem}
\usepackage{ulem}

\makeatletter

\providecolor{lyxadded}{rgb}{0,0,1}
\providecolor{lyxdeleted}{rgb}{1,0,0}
\DeclareRobustCommand{\mklyxadded}[1]{\textcolor{lyxadded}\bgroup#1\egroup}
\DeclareRobustCommand{\mklyxdeleted}[1]{\textcolor{lyxdeleted}\bgroup\mklyxsout{#1}\egroup}
\DeclareRobustCommand{\mklyxsout}[1]{\ifx\\#1\else\sout{#1}\fi}

\usepackage{amsmath,amssymb}
\usepackage[level=0]{wgroup_message}  
\usepackage{color}  
\usepackage{graphicx,psfrag,cite,subfigure}

\usepackage{flushend}
\author{
Wangqian Chen,
Junting~Chen and
Shuguang~Cui
\thanks{This work was supported in part by the National Science Foundation of China (NSFC)
under Grant 62293482, in part by Basic Research Project under Grant HZQB-KCZYZ-2021067 of
Hetao Shenzhen-HK S\&T Cooperation Zone, in part by NSFC Grant 62171398,
in part by Shenzhen Science and Technology Program under Grant JCYJ20220530143804010,
Grant KJZD20230923115104009, and Grant KQTD20200909114730003,
in part by Guangdong Basic and Applied Basic Research Foundation 2024A1515011206,
in part by Guangdong Research Grant 2019QN01X895,
in part by the Guangdong Provincial Key Laboratory of Future Networks of Intelligence
under Grant 2022B1212010001 and Guangdong-Hong Kong-Macao Joint Laboratory for Millimeter-Wave
and Terahertz under Grant 2023B1212120002, in part by the National Key R\&D Program of China
under Grant 2018YFB1800800, and in part by the Key Area R\&D Program of Guangdong Province
under Grant 2018B030338001.
}
\thanks{W.~Chen, J.~Chen and S.~Cui are with the School of Science and Engineering,
the Shenzhen Future Network of Intelligence Institute (FNii-Shenzhen),
and the Guangdong Provincial Key Laboratory of Future Networks of Intelligence,
The Chinese University of Hong Kong, Shenzhen, Guangdong 518172, China
(email: wangqianchen@link.cuhk.edu.cn; juntingc@cuhk.edu.cn; shuguangcui@cuhk.edu.cn).
}
}
\usepackage[acronym]{glossaries}
\newcommand{\newac}{\newacronym}
\newcommand{\ac}{\gls}
\newcommand{\Ac}{\Gls}
\newcommand{\acpl}{\glspl}

\makeglossaries
\newac{speb}{SPEB}{square position error bound}
\newac[plural=EFIMs,firstplural=Fisher information matrices (EFIMs)]{efim}{EFIM}{Fisher information matrix}
\newac{ne}{NE}{Nash equilibrium}
\newac{mse}{MSE}{mean squared error}
\newac{toa}{TOA}{time-of-arrival}
\newac{snr}{SNR}{signal-to-noise ratio}
\newac{lan}{LAN}{local area network}
\newac{psd}{PSD}{positive semidefinite}
\newac{pd}{PD}{positive definite}
\newac{wrt}{w.r.t.}{with respect to}
\newac{lhs}{L.H.S.}{left hand side}
\newac{wp1}{w.p.1}{with probability 1}
\newac{kkt}{KKT}{Karush-Kuhn-Tucker}
\newac{wlog}{w.l.o.g.}{without loss of generality}
\newac{mle}{MLE}{maximum likelihood estimation}
\newac{gps}{GPS}{global positioning system}
\newac{rssi}{RSSI}{received signal strength indicator}
\newac{mimo}{MIMO}{multiple-input multiple-output}
\newac{csi}{CSI}{channel state information}
\newac{fdd}{FDD}{frequency division duplexing}
\newac{ms}{MS}{mobile station}
\newac{bs}{BS}{base station}
\newac{d2d}{D2D}{device-to-device}
\newac{slnr}{SLNR}{signal-to-interference-leakage-and-noise-ratio}
\newac{ula}{ULA}{uniform linear antenna array}
\newac{pas}{PAS}{power angular spectrum}
\newac{mmse}{MMSE}{minimum mean square error}
\newac{zf}{ZF}{zero-forcing}
\newac{rzf}{RZF}{regularized zero-forcing}
\newac{as}{AS}{angular spread}
\newac{aod}{AOD}{angle of departure}
\newac{iid}{i.i.d.}{independent and identically distributed} 
\newac{sinr}{SINR}{signal-to-interference-and-noise ratio}
\newac{tdd}{TDD}{time-division duplex}
\newac{rvq}{RVQ}{random vector quantization}
\newac{rhs}{R.H.S.}{right hand side}
\newac{mrc}{MRC}{maximum ratio combining}
\newac{cdf}{CDF}{cumulative distribution function}
\newac{a.s.}{a.s.}{almost surely}
\newac{los}{LOS}{line-of-sight}
\newac{jsdm}{JSDM}{joint spatial division and multiplexing}
\newac{map}{MAP}{maximum a posteriori}
\newac{klt}{KLT}{Karhunen-Lo\`eve Transform}
\newac{lbe}{LBE}{link bargaining equilibrium}
\newac{se}{SE}{Stackelberg equilibrium}
\newac{uav}{UAV}{unmanned aerial vehicle}
\newac{nlos}{NLOS}{non-line-of-sight}
\newac{pdf}{PDF}{probability density function}
\newac{em}{EM}{expectation-maximization}
\newac{knn}{KNN}{$k$-nearest neighbor}
\newac{svd}{SVD}{singular value decomposition}
\newac{nmf}{NMF}{non-negative matrix factorization}
\newac{umf}{UMF}{unimodality-constrained matrix factorization}
\newac{rmse}{RMSE}{rooted mean squared error}
\newac{olos}{OLOS}{obstructed line-of-sight}
\newac{mmw}{mmW}{millimeter wave}
\newac{ber}{BER}{bit error rate}
\newac{rss}{RSS}{received signal strength}
\newac{lp}{LP}{linear program}
\newac{ufw}{U-FW}{unimodal Frank-Wolfe}
\newac{utf}{UTF}{unimodality-constrained tensor factorization}
\newac{fw}{FW}{Frank-Wolfe}
\newac{iot}{IoT}{Internet-of-Things}
\newac{mae}{MAE}{mean absolute error}
\newac{crb}{CRB}{Cram\'er-Rao bound}
\newac{aoa}{AoA}{angle of arrival}
\newac{wcl}{WCL}{weighted centroid localization}

\usepackage{enumitem}
\usepackage{algorithmic}
\usepackage{multirow}
\usepackage{makecell}
\usepackage{color}
\newac{dl}{DL}{deep learning}
\newac{rt}{RT}{ray tracing}
\newac{nn}{NN}{neural network}
\newac{tx}{TX}{transmitter}
\newac{rx}{RX}{receiver}
\newac{cc}{CC}{channel charting}
\newac{ue}{UE}{user equipment}
\newac{rnn}{RNN}{recurrent neural network}
\newac{cnn}{CNN}{convolutional neural network}
\newac{vae}{VAE}{variational autoencoder}

\makeatother

\theoremstyle{plain}
\newtheorem{thm}{\protect\theoremname}
\newtheorem{prop}{\protect\propositionname}
\usepackage{babel}
\providecommand{\propositionname}{Proposition}
\providecommand{\theoremname}{Theorem}

\begin{document}
\title{Self-Localizing MIMO Beam Mapping with Continuously Evolving Channel
Memory}
\maketitle
\begin{abstract}
Machine learning has greatly advanced data-driven channel modeling
and resource optimization. However, most existing methods require
accurately location-labeled datasets, which are costly to collect
and maintain in dynamic environments. This paper develops a self-localizing
\ac{mimo} beam map framework that constructs a hierarchical wireless
memory from highly sparse \ac{csi} measurements without explicit
location labels. To reduce acquisition and processing overhead, we
use beam-domain \ac{rss} as compact inputs and theoretically show
that they enable asymptotically unbiased spatial signature estimation.
A dual-scale extractor captures intra-snapshot angular dependencies
and inter-sample correlations for incomplete observations, and a hybrid
temporal encoder is designed to consolidate recent CSI into stable
short-term context for physical anchor inference. The inferred anchors
spatially index a physically structured radio map embedding that stores
long-term channel knowledge, which conditions a diffusion decoder
for location-consistent full CSI reconstruction. Such a radio map
embedding provides a persistent wireless knowledge representation
that can be continuously updated and reused without full CSI acquisition.
Experiments show that the proposed framework improves physical-anchor
recovery accuracy by over 30\% under sparse measurements and achieves
more than 20\% channel-capacity gain in \ac{nlos} beam tracking over
Kalman-filter-based methods.
\end{abstract}

\begin{IEEEkeywords}
Blind radio map construction, wireless knowledge memory, \ac{csi}
generation, beam tracking
\end{IEEEkeywords}

\section{Introduction}

\IEEEPARstart{M}{assive} \ac{mimo} has emerged as a cornerstone
technology for 5G and beyond due to its capabilities in spatial multiplexing,
beamforming gain, and interference mitigation. However, to achieve
the full benefit of massive \ac{mimo}, high-dimensional \ac{csi}
is essential, which incurs significant channel training overhead.
As wireless networks become denser and more dynamic, acquiring accurate
\ac{csi} to maintain reliable connectivity and performance becomes
increasingly challenging. To address this issue, channel knowledge
databases incorporating radio maps have emerged as a promising solution
to capture environment-aware channel characteristics \cite{ZengChenXu:J24,PenKalTao:J25}.
In particular, radio maps enable efficient target localization and
beam alignment without exhaustive channel training \cite{HeDongWang:J23,XueJiShaodan:J24},
and have been leveraged for hybrid beamforming and interference coordination
to improve energy efficiency \cite{WuZengJin:J24,ChikhaMasson:J24}.

The fundamental challenge of radio map construction is the demand
for a large volume of accurately \emph{location-labeled} data. Existing
approaches, including statistical modeling \cite{HuZha:J20,DalRos:J22},
interpolation techniques \cite{PhiTonSic:C12,ZhaWan:J22}, tensor
completion \cite{SunChe:J24,ZhaFu:J20}, and deep learning models
\cite{ZhaWij:J23,WangqianChen:J24,JiaCheng:J25}, all face the implementation
challenge of requiring location-labeled \ac{csi} data during the
radio map construction phase. Some recent advances \cite{WangTaoNan:J25,LuoLiPeng:J25}
in generative models attempt to synthesize radio maps directly from
city maps; however, they still require real radio maps with locations
for training. Even if this challenge is addressed during the training
and construction phase, the constructed radio map may still be unreliable,
since minor inaccuracies in location labels or environment dynamics
can affect the radio map accuracy. Moreover, building a ground-truth
reference system is costly, and the dynamic nature of wireless environments
necessitates repeated measurements whenever physical conditions change.
Finally, growing concerns over user location privacy limit the availability
of large-scale location-labeled \ac{csi} data in crowd-sourced settings.

\Ac{rt} methods can bypass the need of location-labeled data for
radio map construction through simulated signal propagation, including
reflection, diffraction, and scattering \cite{HeAiKe:J19,SugaSasakiOsawa:J21}.
While \ac{rt} provides detailed spatial information about \ac{csi},
it incurs prohibitive computational and memory overhead due to the
complexity of modeling intricate propagation mechanisms. In addition,
it requires accurate environmental geometry and material electromagnetic
properties, which may be difficult to obtain and maintain in the network.

\Ac{cc} alleviates the reliance on location-labeled \ac{csi} data
by learning low-dimensional embeddings that preserve local geometric
structures \cite{kazHan:J23,StuMed:J18}. However, \ac{cc} only captures
the relative spatial relationships in a latent space, and requires
additional calibration to align with the physical space. Several works
employed affine transformations from a small set of location-labeled
\ac{csi} samples \cite{ZhangSaad:C21,KarmanovZan:C21,StahlkeYammine:J23},
which still require true location labels. The work \cite{TanerPal:J25}
leveraged access points instead of \ac{ue} positions by encouraging
\acpl{ue} to be closer to the access point from which they receive
stronger signals. But this approach is effective only in \ac{los}
scenarios, and degrades significantly in \ac{nlos} environments with
rich multipath effects.

This paper develops a self-localizing MIMO beam map framework with
hierarchical wireless channel memory using very sparse \ac{csi} measurements
\emph{without} explicit location labels. Unlike conventional label-dependent
approaches, we propose to recover intermediate \ac{ue} locations
as physical anchors that align \ac{csi} observations with a learnable
radio map embedding, which acts as a spatial channel knowledge memory
for \ac{csi} generation and downstream beam tracking. While some
earlier attempt \cite{ZhengChen:J25} employed a model-based Kalman
filter to estimate the \ac{ue} locations, it relies on Gaussian assumptions
and Markovian mobility models, which limits its application in realistic
environments where multipath effects lead to non-Gaussian behavior
and mobility can be non-Markovian.

Specifically, the key challenges to be addressed are threefold. First,
sparse CSI measurements provide only coarse and noisy location cues
that may not align with the physical environment. To address this
challenge, we introduce a self-localizing and physically indexed radio
map embedding that continuously assimilates unlabelled historical
\ac{csi} observations into location-dependent channel representations,
and conditions a diffusion decoder for full \ac{csi} generation with
physical consistency. Second, due to strict pilot, signaling and interface
constraints, full-dimensional complex \ac{csi} data is often unavailable
or costly to transmit. We theoretically show that the \ac{rss} of
MIMO beams provide an effective spatial signature for physical anchoring.
We then develop a dual-scale feature extractor that jointly captures
intra-snapshot angular dependencies via self-attention and inter-sample
correlations with multi-scale convolutions, thus enhancing data efficiency
and robustness. Third, location inference can be unstable in dynamic
environments, where complex mobility and short-term multipath fluctuations
obscure the underlying spatial evolution and cause error accumulation.
To obtain stable physical anchors, we develop a temporal encoder that
aggregates historical information context through bidirectional recurrent
modeling, refines local dynamics with temporal convolutions, and truncates
unreliable states. By exploiting historical context rather than individual
snapshots, it suppresses channel variations and enables robust inference
over time.

The novelty and contributions are summarized as follows.
\begin{itemize}
\item We formulate self-localizing radio map construction as a physically
anchored generative problem and develop a radio-map-embedded architecture.
Without explicit location labels, the framework infers physical anchors
to align sparse CSI observations with a physically indexed radio map
memory, thereby jointly learning physical anchoring, long-term radio
map learning, and location-consistent full CSI reconstruction within
a unified framework.
\item We provide a theoretical basis for using \ac{rss} sequences as input
signatures for radio map construction. We show that the RSS of MIMO
beams yields an asymptotically unbiased \ac{aod} estimate at high
signal-to-noise ratio (SNR) for physical anchoring. This motivates
an RSS-based structure that avoids modeling of high-dimensional complex
CSI, simplifying network feature processing while retaining location-relevant
information.
\item We develop a hierarchical memory scheme that converts CSI measurements
into persistent and evolving radio map memory. A history-aware temporal
encoder consolidates recent CSI observations into stable short-term
representations, which are accumulated into a incrementally updated
long-term radio map embedding.
\item We conduct experiments to show that the proposed model improves physical
anchor recovery accuracy by over 30\% under very sparse measurements
and achieves more than 20\% average channel capacity gain in \ac{nlos}
beam tracking compared with Kalman-filter-based baselines.
\end{itemize}

The rest of the paper is organized as follows. Section II reviews
the system model and the RSS-based feature motivation. Section III
develops the physically anchored generative radio-map learning architecture.
Section IV presents experimental results, and Section V concludes
the paper.

{\em Notations:} $(\cdot)^{\text{T}}$ and $(\cdot)^{\text{*}}$
represent the transpose and Hermitian transpose operations, respectively.
$\left|\cdot\right|$ denotes the absolute value and $\left\Vert \cdot\right\Vert ^{2}$
represents the $L_{2}$ norm. \ensuremath{\odot} denotes the element-wise
product, and $\mathcal{N}(\cdot,\cdot)$ denotes a Gaussian distribution.

\section{Location Recoverable CSI Feature}

\subsection{System Model}

Consider a massive \ac{mimo} communication system with $N_{b}$ \acpl{bs}
and multiple mobile \acpl{ue}, where each \ac{bs}, treated as a
\ac{tx}, is equipped with $N_{t}$ antennas, and all \acpl{ue},
treated as \acpl{rx}, are single-antenna devices. Denote a link $\tilde{\mathbf{p}}=(\mathbf{p}_{\mathrm{t}},\mathbf{p}_{\mathrm{r}})\in\mathbb{R}^{6}$
with the positions $\mathbf{p}_{\mathrm{t}},\mathbf{p}_{\mathrm{r}}\in\mathbb{R}^{3}$
of the \ac{tx} and the \ac{rx}, respectively. 

The narrow-band \ac{mimo} channel of $\tilde{\mathbf{p}}$ can be
modeled as
\begin{equation}
\mathbf{h}(\tilde{\mathbf{p}})=\sum^{L_{p}}_{l=0}\beta_{l}(\tilde{\mathbf{p}})\mathbf{a}(\phi_{l}(\tilde{\mathbf{p}}))\label{eq:channel_vector}
\end{equation}
where $\beta_{l}(\tilde{\mathbf{p}})$ and $\phi_{l}(\tilde{\mathbf{p}})$
denote the gain and \ac{aod} of the $l$th path, and $\mathbf{a}(\phi_{l})$
is the array steering vector.

A codebook-based approach at the \ac{bs} is applied to construct
MIMO beams for channel measurement. Let $\mathbf{w}_{j}$ denote the
$j$th beamforming vector from a codebook $\mathcal{W}$, satisfying
the power constraint $\|\mathbf{w}_{j}\|^{2}=1$. The received signal
of the MIMO beam under the beamforming vector $\mathbf{w}_{j}$ is
given by 
\begin{equation}
\rho(\tilde{\mathbf{p}},\mathbf{w}_{j})=\mathbf{h}^{*}(\tilde{\mathbf{p}})\mathbf{w}_{j}+n_{j}\label{eq:channel_gain_model}
\end{equation}
where $n_{j}$ denotes the measurement noise. The beamforming vectors
$\mathbf{w}_{j}$ are designed according to the antenna geometry such
that the beam energy concentrates around one specific direction. An
example under \ac{ula} is to adopt the discrete Fourier transform
(DFT) codebook.

Here, accurate \ac{ue} location is not assumed to be available. Our
goal is to jointly infer the location and CSI from sparse measurements.
We propose to use a radio map to structure the historical CSI data.
Thus the fundamental issues are to determine what CSI feature the
system should memorize, and how to construct the radio map from pure
CSI measurements without explicit location labels.

\subsection{Location Recoverability from the RSS of the MIMO Beams}

We first establish the theoretical foundation of recovering the UE
location from a sequence of CSI measurements. Here, we consider using
the \ac{rss} of the \ac{mimo} beams as the location signature of
the \ac{ue} due to its low acquisition overhead.

Given the codebook $\mathcal{W}$, each beam induces a deterministic
beam pattern with respect to the \ac{aod}. For a fixed link $\tilde{\mathbf{p}}$,
we assume that there exists a dominant propagation direction with
\ac{aod} $\phi_{D}(\tilde{\mathbf{p}})$. It may follow a Gaussian
distribution centered at the geometric azimuth angle $\varphi(\mathbf{p}_{\mathrm{t}},\mathbf{p}_{\mathrm{r}})$
from the BS to the \ac{ue}, i.e., $\phi_{D}(\tilde{\mathbf{p}})\sim\mathcal{N}(\varphi(\mathbf{p}_{\mathrm{t}},\mathbf{p}_{\mathrm{r}}),\sigma^{2}_{\varphi})$,
where $\sigma^{2}_{\varphi}$ captures the angular uncertainty due
to local scattering. We then find that there exists a maximum likelihood
estimate (MLE) of $\phi_{D}(\tilde{\mathbf{p}})$ from the RSS measurements
of the MIMO beams, and such an estimator is approximately unbiased
with bounded variance as characterized in the following theorem.

Let $\mathbf{g}=[|\rho(\tilde{\mathbf{p}},\mathbf{w}_{j})|^{2}]_{j\in\mathcal{W}}\in\mathbb{R}^{|\mathcal{W}|}$
be the RSS vector of $\tilde{\mathbf{p}}$, and define its normalized
form as $\widetilde{\mathbf{g}}\triangleq\mathbf{g}/(\mathbf{1}^{T}\mathbf{g})$,
where $|\mathcal{W}|$ is the number of beamforming vectors. Let the
deterministic beam pattern be $b_{j}(\phi_{D})\triangleq|\mathbf{a}^{*}(\phi_{D}(\tilde{\mathbf{p}}))\mathbf{w}_{j}|^{2}$,
and define $\mathbf{b}(\phi_{D})=[b_{j}(\phi_{D})]_{j\in\mathcal{W}}\in\mathbb{R}^{|\mathcal{W}|}$.
The corresponding normalized vector is then given by $\mathbf{f}(\phi_{D})\triangleq\mathbf{\mathbf{b}}(\phi_{D})/(\mathbf{1}^{T}\mathbf{\mathbf{\mathbf{b}}}(\phi_{D}))$.
\begin{thm}
\label{thm:Proposed1} (RSS as an efficient Spatial Signature) The
MLE of the \ac{aod} $\phi_{D}(\tilde{\mathbf{p}})$ is asymptotically
unbiased and follows a Gaussian distribution
\begin{equation}
\hat{\phi}_{D}(\tilde{\mathbf{p}})\sim\mathcal{N}(\phi_{D}(\tilde{\mathbf{p}}),(\mathbf{f}'(\phi_{D})^{T}\Sigma^{-1}_{\eta}\mathbf{f}'(\phi_{D}))^{-1})\label{eq:AOD_model-1}
\end{equation}
at high SNR, where $\mathbf{f}'(\phi)\triangleq\partial\mathbf{f}(\phi)/\partial\phi$
denotes the derivative of $\mathbf{f}(\phi)$ with respect to $\phi$,
and $\Sigma_{\eta}$ is the covariance matrix of the normalized RSS
perturbation.
\end{thm}
\begin{proof}
See Appendix~\ref{app:Theom proof}.
\end{proof}
Theorem~\ref{thm:Proposed1} reveals that the RSS of the MIMO beams
already contains sufficient information on the \ac{ue} direction,
and can therefore serve as a compact spatial signature. Once we have
found an unbiased \ac{aod} estimator, existing results in the literature
show that the mobile trajectory can be recovered from sequences of
AOD observations.
\begin{prop}
\label{lem:The-Cramer-Rao} (Trajectory Recoverability from \ac{aod}
sequences \cite{XingChen:J25}) Consider a \ac{ue} following a linear
mobility model over $L$ time slots, i.e., $\mathbf{p}^{l}_{\mathrm{r}}=\mathbf{p}^{0}_{\mathrm{r}}+l\boldsymbol{v}$
for $l=1,...,L$, where $\mathbf{p}^{0}_{\mathrm{r}}$ and $\boldsymbol{v}$
denote the initial position and velocity, respectively. Given a sequence
of AOD observations $\{\phi_{D}(\mathbf{p}_{\mathrm{t}},\mathbf{p}^{l}_{\mathrm{r}})\}^{L}_{l=1}$
related to $\mathbf{p}^{l}_{\mathrm{r}}$, the Cramer-Rao Lower Bound
(CRLB) of $\mathbf{p}^{0}_{\mathrm{r}}$ scales asymptotically as
$\mathcal{O}(1/(SNR\cdot L\cdot N^{3}_{t}))$, while that of \textup{$\boldsymbol{v}$}
scales as $\mathcal{O}(1/(SNR\cdot L^{3}\cdot N^{3}_{t}))$.
\end{prop}
Proposition~\ref{lem:The-Cramer-Rao} guarantees the recoverability
for rectilinear mobility from AOD sequences. It shows that the trajectory
is fully recoverable with diminishing CRLB as $L$ goes to infinity,
under arbitrarily poor SNR. While Proposition~\ref{lem:The-Cramer-Rao}
merely studies rectilinear mobility, it provides a strong support
for trajectory recoverability in practice because a real-life trajectory
usually consists of a number of linear segments.

Combining Theorem~\ref{thm:Proposed1} and Proposition~\ref{lem:The-Cramer-Rao},
it suffices to use RSS measurements as features for learning hidden
location from CSI. This simplifies the network design from processing
complex-valued CSI, and potentially leads to a simpler network structure
that is easier to train.

\begin{figure}
\centering\includegraphics[scale=0.48]{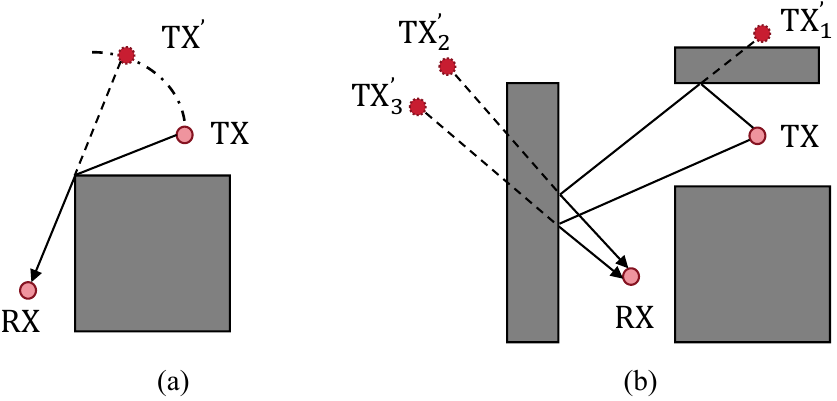}\caption{Virtual-source interpretation of reflected and diffracted paths: (a)
Equivalent virtual TX for diffraction; (b) Mirror TX lattice for reflections.}
\label{fig:Recoverability under Blockage and Reflection}
\end{figure}

\subsubsection*{Remark (Recoverability under blockage and reflection)}

For the case where it is reflection and diffraction, one can model
the propagation paths shown in Fig.~\ref{fig:Recoverability under Blockage and Reflection}.
For example, a diffraction path is modeled by a virtual TX whose line
to the RX passes through the diffraction point while preserving the
propagation distance. Similarly, each specular reflection is modeled
by mirroring the TX with respect to the reflecting surface. As a result,
the AOD of the reflective or diffracted path can be modeled using
a Gaussian distribution centered at the azimuth angle $\varphi(\mathbf{p}_{\mathrm{t}^{'}},\mathbf{p}_{\mathrm{r}})$
of the virtual TX at $\mathbf{p}_{\mathrm{t}^{'}}$. Theorem~\ref{thm:Proposed1}
and Proposition~\ref{lem:The-Cramer-Rao} then imply that the RSS
of these paths are also efficient spatial signatures for the UE location.
Note that the topology of the virtual TXs only depends on the environment,
not on the UE location, and hence, is fixed.

While we do not assume the positions of the virtual TXs, our goal
is to develop deep learning models to implicitly learn the topology
information of these TXs as a consequence of the propagation environment
using the RSS measurements. Theorem~\ref{thm:Proposed1} and Proposition~\ref{lem:The-Cramer-Rao}
guarantee the location recoverability from sufficient RSS measurements.

\begin{figure*}
\centering\includegraphics[scale=0.95]{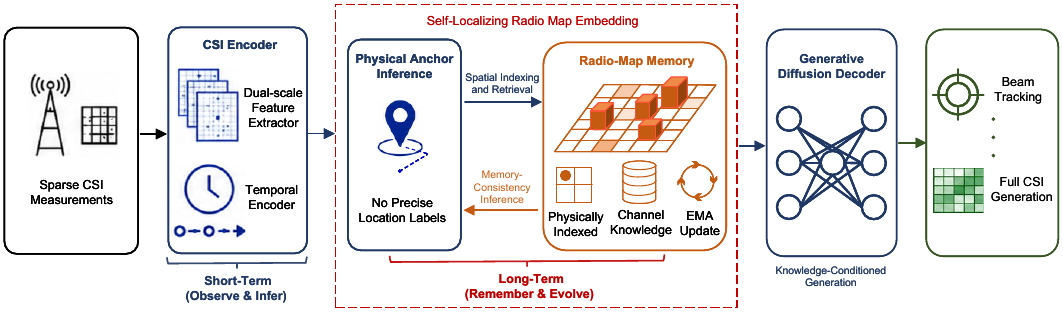}\caption{Self-localizing MIMO beam map framework with a hierarchical and evolving
channel memory. Sparse CSI observations are consolidated by a short-term
temporal encoder, aligned through physical anchors, and accumulated
into a physically structured long-term radio map memory. The retrieved
memory then conditions the diffusion decoder for location-consistent
full CSI reconstruction and downstream beam tracking.}
\label{fig:RadioMap_Arc}
\end{figure*}

\subsection{Self-Localizing MIMO Beam Map Construction Problem}

Let $\mathring{\mathcal{G}}_{L}=\{\boldsymbol{\mathring{g}}^{0}_{\mathrm{r}},\boldsymbol{\mathring{g}}^{1}_{\mathrm{r}},...,\boldsymbol{\mathring{g}}^{L}_{\mathrm{r}}\}$
be the sequence of sparse observations collected along an unknown
trajectory $\mathcal{P}_{L}=\{\mathbf{p}^{0}_{\mathrm{r}},\mathbf{p}^{1}_{\mathrm{r}},...,\mathbf{p}^{L}_{\mathrm{r}}\}$,
where each $\boldsymbol{\mathring{g}}^{l}_{\mathrm{r}}$ contains
only a small subset of the full \ac{csi} vector $\boldsymbol{g}^{l}_{\mathrm{r}}$.
Our goal is to jointly recover the trajectory $\mathcal{P}_{L}$ and
build a radio map model $\mathcal{M}=\{(\mathbf{p}_{\mathrm{r}},\boldsymbol{g}_{\mathrm{r}}):\mathbf{p}_{\mathrm{r}}\in\mathcal{X}\}$,
using only the sparse observations $\mathring{\mathcal{G}}_{L}$ without
access to the true location labels $\mathcal{P}_{L}$, where $\mathcal{X}$
denotes the set of discretized grid cells in the region of interest.
To this end, we formulate a joint maximum log-likelihood problem:

\begin{equation}
\begin{aligned}\mathrm{\underset{\mathcal{G}_{\mathit{L}},\mathcal{P}_{\mathit{L}}}{\textrm{maximize}}}\hspace{0.3cm} & \log p(\mathcal{G}_{L},\mathcal{P}_{L};\mathcal{M}|\mathring{\mathcal{G}}_{L})\\
\mathrm{\textrm{subject\hspace{0.1cm}to}\hspace{0.3cm}} & \hspace{0.05cm}\mathbf{p}^{l}_{\mathrm{r}}\in\mathcal{X},\hspace{0.3cm}l=0,1,...L.
\end{aligned}
\label{eq:Problem_formulation}
\end{equation}
where $p(\mathcal{G}_{L},\mathcal{P}_{L};\mathcal{M}|\mathring{\mathcal{G}}_{L})$
denotes the joint distribution of \ac{ue} locations and their CSI
conditioned on the observations $\mathring{\mathcal{G}}_{L}$.

The joint objective formalizes the self-localizing radio-map construction
problem but is not directly tractable due to the strong coupling between
the unknown locations and the radio map. We therefore decompose it
into learnable conditional relations that can explicitly guide the
network design.

\section{Hierarchical Channel Memory Framework for Self-Localizing MIMO Beam
Mapping}

We develop a radio-map-embedded framework to solve the problem in
(\ref{eq:Problem_formulation}). The core idea is to factorize the
joint probability $p(\mathcal{G}_{L},\mathcal{P}_{L};\mathcal{M}|\mathring{\mathcal{G}}_{L})$
using a recurrent probabilistic structure, and map each component
to a specific neural network module.

We introduce a latent state $\boldsymbol{S}^{l}$ that summarizes
historical information, which evolves recursively as
\begin{equation}
\begin{aligned}\boldsymbol{S}^{l}= & f_{\theta}(\mathbf{p}^{l-1}_{\mathrm{r}},\boldsymbol{\mathring{g}}^{l-1}_{\mathrm{r}},\boldsymbol{S}^{l-1})\end{aligned}
\label{eq:joint probability-1-1}
\end{equation}
where $f_{\theta}(\cdot)$ is a learnable transition function. Here,
$\boldsymbol{S}^{l}$ represents a learned mobility context rather
than a fixed Markovian mobility model. It can be implemented over
a sliding window to exploit short-term temporal channel correlations.

Based on $\boldsymbol{S}^{l}$, the joint distribution admits a recurrent
form:
\begin{equation}
\begin{aligned}p(\mathcal{G}_{L},\mathcal{P}_{L};\mathcal{M}|\mathring{\mathcal{G}}_{L}) & =p(\mathbf{p}^{0}_{\mathrm{r}}|\boldsymbol{\mathring{g}}^{0}_{\mathrm{r}};\mathcal{M})\prod^{L}_{l=1}p(\mathbf{p}^{l}_{\mathrm{r}}|\boldsymbol{S}^{l},\boldsymbol{\mathring{g}}^{l}_{\mathrm{r}};\mathcal{M})\\
 & \times\prod^{L}_{l=0}p(\boldsymbol{g}^{l}_{\mathrm{r}}|\mathbf{p}^{l}_{\mathrm{r}};\mathcal{M}).
\end{aligned}
\label{eq:joint probability-2}
\end{equation}
Accordingly, the MLE objective in (\ref{eq:Problem_formulation})
can be written as
\begin{equation}
\begin{aligned}\mathrm{\underset{\mathcal{G}_{\mathit{L}},\mathcal{P}_{\mathit{L}}}{\textrm{maximize}}}\hspace{0.2cm} & \sum^{L}_{l=1}\log p(\mathbf{p}^{l}_{\mathrm{r}}|\boldsymbol{S}^{l},\boldsymbol{\mathring{g}}^{l}_{\mathrm{r}};\mathcal{M})+\log p(\mathbf{p}^{0}_{\mathrm{r}}|\boldsymbol{\mathring{g}}^{0}_{\mathrm{r}};\mathcal{M})\\
 & +\sum^{L}_{l=0}\log p(\boldsymbol{g}^{l}_{\mathrm{r}}|\mathbf{p}^{l}_{\mathrm{r}};\mathcal{M})\\
\mathrm{\textrm{subject\hspace{0.1cm}to}\hspace{0.2cm}} & \hspace{0.05cm}\mathbf{p}^{l}_{\mathrm{r}}\in\mathcal{X},\hspace{0.3cm}l=0,1,...L.
\end{aligned}
\label{eq:Problem_formulation2}
\end{equation}

This factorization identifies the functional roles of the network.
The first two terms correspond to self-localizing physical anchors
from sparse observations and historical context, while the final term
models location-consistent full CSI generation. The radio map links
these components by learning the correspondence between physical space
and channel characteristics. Although this formulation resembles a
latent-variable model, such as a variational autoencoder (VAE), its
direct implementation remains challenging. First, severe sparsity
and noise can obscure the spatial-temporal signatures required for
reliable physical anchoring. Second, standard VAE regularization does
not impose an explicit correspondence between latent variables and
the radio map, while discrete map grids make end-to-end learning non-differentiable.
Consequently, maximum-likelihood learning alone cannot ensure the
physical alignment required for radio map construction.

To address these issues, we develop a self-localizing MIMO beam map
framework illustrated in Fig.~\ref{fig:RadioMap_Arc}. It transforms
transient sparse observations into persistent, physically indexed
channel representations through the joint learning of physical anchors
and radio map memory. First, sparse observations are encoded into
robust channel representations, where historical measurements are
aggregated to form short-term context for reliable physical anchor
inference. These anchors provide the spatial indexing mechanism to
retrieve the long-term radio-map memory, while memory consistency
provides spatial feedback to refine the inferred anchors. The retrieved
memory representation subsequently conditions the generative decoder
to reconstruct location-consistent full CSI. Unlike vector-quantized
VAEs (VQ-VAEs) \cite{VQ-VAE:J17}, whose codebook entries have no
inherent physical interpretation, the proposed memory is explicitly
associated with physical locations. Each entry represents the accumulated
channel representation of a specific spatial region and can be continuously
updated with new observations. Therefore, the radio map is not a static
database built after localization, but an evolving wireless memory
that jointly learns spatial correspondence and channel knowledge from
sparse measurements.

In summary, the proposed framework includes three closely connected
designs. The short-term temporal encoder stabilizes inference from
sparse observations; the coupled physical-anchor and radio-map module
converts unlabeled CSI streams into physically indexed long-term memory;
and the diffusion decoder exploits the channel memory for full CSI
reconstruction and beam tracking. The following subsections present
the detailed implementation of these components.

\begin{figure*}[!t]
\centering \includegraphics[scale=0.57]{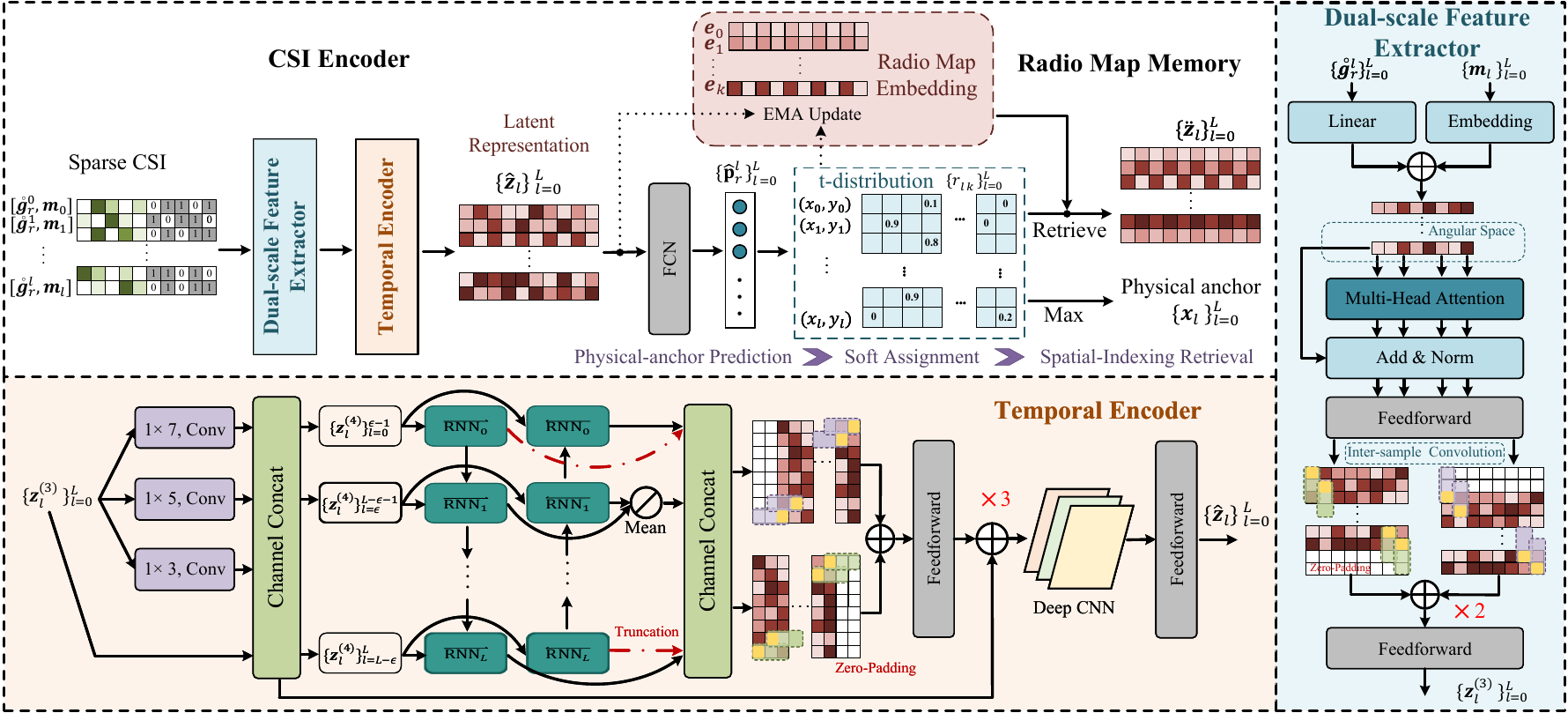}\caption{Detailed architecture of the hierarchical and evolving channel memory
framework for self-localizing MIMO beam map construction. A dual-scale
extractor derives robust spatial signatures from sparse CSI, a temporal
encoder consolidates recent observations into stable short-term context
for physical anchor inference, and a physically indexed radio map
memory accumulates the representations as reusable long-term, location-dependent
channel knowledge.}
\label{fig:network-architecture-encoder}
\end{figure*}

\subsection{Dual-scale Feature Extraction}

The first challenge is to obtain reliable spatial signatures from
sparse observations. Since only a small subset of CSI measurements
is available, the observed CSI pattern can be incomplete and ambiguous.
Moreover, missing entries should not be treated as zero-power measurements,
as this may introduce artificial patterns and lead to unstable physical
anchors.

To address this issue, we design a mask-aware dual-scale feature extractor
that exploits CSI correlations at two scales. First, within each CSI
snapshot, neighboring beams are correlated due to beam leakage, angular
spread, and multipath coupling. Such angular dependencies help infer
the structure of partially observed beam patterns. Second, adjacent
CSI samples exhibit smooth variations over time, which provides inter-sample
context for denoising and pattern completion. The feature extractor
thus jointly captures intra-snapshot angular dependencies and inter-sample
spatio-temporal correlations.

Let $\mathbf{G},\mathbf{M}\in\mathbb{R}^{L\times N_{b}|\mathcal{W}|}$
be the stacked observations $\boldsymbol{\mathring{g}}^{l}_{\mathrm{r}}$
and masking vectors $\boldsymbol{m}^{l}$ over $L$ time steps, where
zero entries in $\boldsymbol{m}^{l}$ indicate unobserved beam entries.
To separate the BS and beam dimensions, the inputs are reshaped into
$\mathbf{G},\mathbf{M}\rightarrow\mathbf{\tilde{G}},\mathbf{\tilde{M}}\in\mathbb{R}^{LN_{b}\times|\mathcal{W}|\times1}$.
We first use a mask-aware embedding layer to jointly encode CSI values
and masking indicators:
\begin{equation}
\begin{aligned}\boldsymbol{\mathbf{z}}^{(0)} & =f_{lin}(\mathbf{\tilde{G}})+f_{emb}(\tilde{\mathbf{M}})\in\mathbb{R}^{LN_{b}\times|\mathcal{W}|\times d_{e}}\end{aligned}
\label{eq:Embedding}
\end{equation}
where $d_{e}$ is the embedding dimension, $f_{lin}(\cdot)$ is a
linear layer, and $f_{emb}(\cdot)$ assigns learnable embeddings to
observed and missing entries. This mask-aware encoding allows the
network to distinguish between observed and unobserved entries.

The angular dependencies within each snapshot are then captured by
self-attention along the beam dimension $|\mathcal{W}|$:
\begin{equation}
\begin{aligned}\boldsymbol{\mathbf{z}}^{(1)} & =f_{trans}(\boldsymbol{\mathbf{z}}^{(0)})\in\mathbb{R}^{LN_{b}\times|\mathcal{W}|\times d_{e}}\end{aligned}
\label{eq:CSI-Feature-Extractor-1}
\end{equation}
where $f_{trans}(\cdot)$ is a Transformer block. This operation aggregates
information across beams and enhances the angular signature even when
only partial beam measurements are observed. The output is reshaped
to $\boldsymbol{\mathbf{z}}^{(1)}\rightarrow\tilde{\boldsymbol{\mathbf{z}}}^{(1)}\in\mathbb{R}^{L\times N_{b}|\mathcal{W}|d_{e}}$.

To exploit inter-sample correlations, double multi-scale convolutions
are applied along the sequence dimension $L$:
\begin{equation}
\begin{aligned}\boldsymbol{\mathbf{z}}^{(2)} & =f_{c1}(\tilde{\boldsymbol{\mathbf{z}}}^{(1)})+\mathrm{flip}(f_{c2}(\mathrm{flip}(\tilde{\boldsymbol{\mathbf{z}}}^{(1)})))\in\mathbb{R}^{L\times N_{b}|\mathcal{W}|d_{e}}\end{aligned}
\label{eq:CSI-Feature-Extractor}
\end{equation}
where $f_{c1/c2}(\cdot)$ denote multi-scale convolutional layers
and $\mathrm{flip}(\cdot)$ reverses the sequence order. This structure
allows double convolutions to aggregate historical and future context
to suppress local noise and compensate for missing entries.

After two stacked attention-convolutional blocks, a fully connected
network (FCN) produces $\boldsymbol{\mathbf{z}}^{(3)}=f_{N1}(\boldsymbol{\mathbf{z}}^{(2)})\in\mathbb{R}^{L\times N_{b}|\mathcal{W}|}$.
This representation mitigates the uncertainty induced by sparse and
noisy measurements, thus providing a more reliable basis for physical-anchor
inference. 

\subsection{Temporal Inference for Physical Anchoring}

Since wireless channels exhibit strong short-term correlation, recurrent
modeling provides a natural mechanism for capturing temporal dependencies.
However, a plain \ac{rnn} may suffer from initial-state bias and
error accumulation due to limited temporal context, while channel
fluctuations may also induce local jitters that do not correspond
to actual UE motion.

To address these issues, we develop a hybrid recurrent-convolutional
temporal context encoder. The recurrent component captures sequential
dependencies with bias truncation to reduce early-sequence errors,
while the convolutional component refines local mobility patterns
and suppresses temporal jitters that do not correspond to actual UE
motion.

We first enrich the input feature using parallel convolutions with
kernel sizes of 3, 5, and 7, and concatenate them as
\begin{equation}
\begin{aligned}\boldsymbol{\boldsymbol{\mathbf{z}}}^{(4)} & =[\boldsymbol{\boldsymbol{\mathbf{z}}}^{(3)},f_{c3}(\boldsymbol{\boldsymbol{\mathbf{z}}}^{(3)}),f_{c4}(\boldsymbol{\boldsymbol{\mathbf{z}}}^{(3)}),f_{c5}(\boldsymbol{\boldsymbol{\mathbf{z}}}^{(3)})]\in\mathbb{R}^{L\times d_{c}}\end{aligned}
\label{eq:Temporal-Context-Extractor}
\end{equation}
where $f_{c3/c4/c5}(\cdot)$ are convolutional layers with different
receptive fields, and $d_{c}$ is the concatenated feature dimension.

The resulting feature sequence is processed by forward and backward
\acpl{rnn} to capture historical and future context:
\begin{equation}
\begin{aligned}\mathbf{R}^{f} & =f_{r1}(\boldsymbol{\boldsymbol{\mathbf{z}}}^{(4)}), & \mathbf{R}^{b}=\mathrm{flip}(f_{r2}(\mathrm{flip}(\boldsymbol{\boldsymbol{\mathbf{z}}}^{(4)})))\in\mathbb{R}^{L\times d_{r}}\end{aligned}
\label{eq:RNN-LAYER}
\end{equation}
where $f_{r1/r2}(\cdot)$ denote \acpl{rnn}, and $d_{r}$ is the
hidden dimension.

However, directly averaging the two recurrent states is unreliable
because of initial-state bias. The forward state has limited past
context at the beginning of the sequence, whereas the backward state
has limited future context near the end. We therefore introduce a
reliability-aware truncation rule:
\begin{equation}
\begin{aligned}\boldsymbol{\boldsymbol{\mathbf{z}}}^{(5)}_{l} & =\begin{cases}
\mathbf{R}^{b}_{l} & l\leq\epsilon\\
(\mathbf{R}^{f}_{l}+\mathbf{R}^{b}_{l})/2 & \epsilon<l<L-\epsilon\\
\mathbf{R}^{f}_{l} & l\geq L-\epsilon
\end{cases}\end{aligned}
\label{eq:RNN-Layer}
\end{equation}
where $\epsilon$ denotes the truncation length. This fusion rule
uses the more reliable backward context near the beginning, the forward
context near the end, and both contexts in the middle, thereby reducing
initial-state inference errors.

To suppress short-term fluctuations and local jitters, the feature
is refined by convolutional layers across dimension $L$:
\begin{equation}
\begin{aligned}\boldsymbol{\boldsymbol{\mathbf{z}}}^{(6)} & =f_{n2}(f_{c6}(\boldsymbol{\boldsymbol{\mathbf{z}}}^{(5)})+\mathrm{flip}(f_{c7}(\mathrm{flip}(\boldsymbol{\boldsymbol{\mathbf{z}}}^{(5)}))))\in\mathbb{R}^{L\times d_{c}}\end{aligned}
\label{eq:RNN-CNN}
\end{equation}
where $f_{c6/c7}(\cdot)$ are convolutional layers to refine the forward
and backward temporal contexts, and $f_{n2}(\cdot)$ is a feedforward
layer to fuse the dual-context features.

After stacking three recurrent-convolutional blocks, the final location-aware
representation is obtained as 
\begin{equation}
\begin{aligned}\hat{\boldsymbol{\mathbf{z}}} & =f_{n3}(f_{cnn}(\boldsymbol{\boldsymbol{\mathbf{z}}}^{(6)}+\boldsymbol{\boldsymbol{\mathbf{z}}}^{(4)}))\in\mathbb{R}^{L\times D}\end{aligned}
\label{eq:DEEP-CNN}
\end{equation}
where $f_{cnn}(\cdot)$ extracts deep feature representations, and
$f_{n3}(\cdot)$ projects them to the target latent dimension $D$.

In summary, this temporal encoder is designed not merely to model
sequence dependence, but to stabilize physical-anchor inference under
complex mobility.

\subsection{Physically Indexed Radio Map Memory Learning}

The purpose of the radio map is not merely to compress channel observations,
but to make the learned representation physically queryable. Standard
VQ-VAE codebooks are formed by feature-space clustering and are not
inherently associated with physical locations. They cannot directly
serve RAN intelligent controller applications that request channel
knowledge for a location or an inferred UE state.

We introduce a physically indexed radio map embedding $\mathcal{E}=[\boldsymbol{e}_{1},...,\boldsymbol{e}_{K}]\in\mathbb{R}^{K\times D}$,
where $\boldsymbol{e}_{k}\in\mathbb{R}^{D}$ denotes the high-level
channel signature associated with the $k$th grid cell in $\mathcal{X}$.
Unlike a VQ-VAE codeword, $\boldsymbol{e}_{k}$ is tied to a physical
grid cell and is used as location-dependent channel knowledge memory.

Each location-aware representation vector $\hat{\boldsymbol{\mathbf{z}}}_{l}$
is first mapped to an intermediate physical anchor,
\begin{equation}
\begin{aligned}\mathbf{\hat{p}}^{l}_{\mathrm{r}} & =f_{NN2}(\hat{\boldsymbol{\mathbf{z}}}_{l})\end{aligned}
\label{eq:FCN-Localization}
\end{equation}
where $f_{NN2}(\cdot)$ is the physical-anchor prediction network.
The anchor is not an externally supplied label; it is the internal
spatial index used to access the knowledge memory. To align the physical
anchor with the radio map, we compute a soft assignment over grid
cells using a Student\textquoteright s t-distribution:
\begin{equation}
\begin{aligned}r_{lk}= & \frac{(1+||\mathbf{\hat{p}}^{l}_{\mathrm{r}}-\boldsymbol{x}_{k}||^{2}/\alpha)^{-\frac{\alpha+1}{2}}}{\sum^{K}_{k'}(1+||\mathbf{\hat{p}}^{l}_{\mathrm{r}}-\boldsymbol{x}_{k'}||^{2}/\alpha)^{-\frac{\alpha+1}{2}}}\end{aligned}
\label{eq:probability estimation}
\end{equation}
where $\boldsymbol{x}_{k}$ is the $k$th grid cell center, and $\alpha$
controls the degree of freedom. The heavy-tailed assignment provides
stable gradients when the early anchor estimates are still unreliable,
while gradually encouraging the inferred anchors to concentrate around
spatially consistent grid cells.

Based on the assignment, the radio map embedding is then retrieved
through spatial indexing:
\begin{equation}
k^{*}=\mathrm{argmax}_{k}|r_{lk}|,\hspace{0.2cm}\ddot{\mathbf{z}}_{l}=\boldsymbol{e}_{k^{*}}.\label{eq:Codebook1}
\end{equation}

Unlike VQ-VAEs based solely on feature similarity, the proposed method
retrieves radio-map embeddings according to spatial association with
the inferred physical anchor. The embedding $\ddot{\mathbf{z}}_{l}$
simultaneously provides a compact channel knowledge representation,
a persistent map entry, and a physical condition for the generative
decoder.

\begin{figure*}[!t]
\centering\includegraphics[scale=0.6]{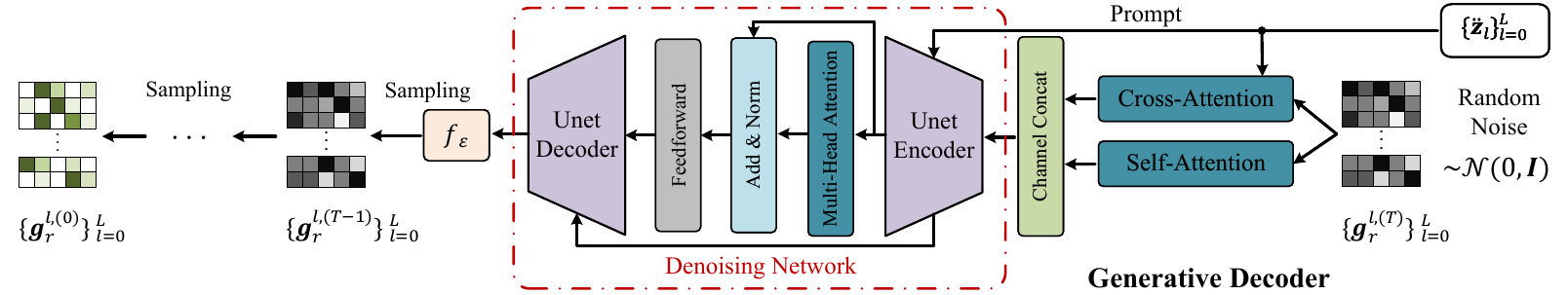}\caption{Knowledge-conditioned diffusion decoder for location-consistent full-CSI
reconstruction.}
\label{fig:network-architecture-1}
\end{figure*}

\subsection{Knowledge-Conditioned Generative CSI Reconstruction}

Some beam-management applications require the full CSI state rather
than only the compact map embedding. Full CSI is high-dimensional
and stochastic under multipath propagation; deterministic decoders
tend to over-smooth the channel, while unconditional generators may
lose spatial consistency. We condition a diffusion decoder on the
retrieved map knowledge.

We adopt a radio-map-conditioned diffusion decoder. Since the retrieved
radio map already encodes the spatial structure and the location-dependent
channel signature, the decoder does not need to learn spatial alignment
from scratch, which reduces the burden on the generative model and
enables the use of standard diffusion backbones with radio-map conditioning.

In the forward diffusion process, the clean CSI feature $\boldsymbol{g}^{l,(0)}_{\mathrm{r}}$
is gradually perturbed by injecting Gaussian noise:
\begin{equation}
\begin{aligned}\boldsymbol{g}^{l,(t)}_{\mathrm{r}} & =\sqrt{\bar{\alpha}_{t}}\boldsymbol{g}^{l,(0)}_{\mathrm{r}}+\sqrt{1-\bar{\alpha}_{t}}\boldsymbol{\varepsilon}_{t},\hspace{0.2cm}\boldsymbol{\varepsilon}_{t}\sim\mathcal{N}(0,\boldsymbol{I})\end{aligned}
\label{eq:Forward-Diffusion-Process-1}
\end{equation}
where $\alpha_{t}=1-\eta_{t}$, $\bar{\alpha}_{t}=\prod^{t}_{i=1}\alpha_{i}$,
$\boldsymbol{\varepsilon}_{t}$ is the injected Gaussian noise, and
$\eta_{t}$ is the variance schedule parameter.

The denoising network predicts the injected noise from the noisy CSI,
the diffusion step, and the radio-map embedding: 
\begin{equation}
\begin{aligned}\boldsymbol{\hat{\varepsilon}}_{t} & =f_{\varepsilon}(\boldsymbol{g}^{l,(t)}_{\mathrm{r}},\ddot{\mathbf{z}}_{l},t).\end{aligned}
\label{eq:Forward-Diffusion-Process-2}
\end{equation}
We implement the denoiser $f_{\varepsilon}(\cdot)$ using an attention-augmented
UNet, where self-attention models dependencies within the CSI vector,
while cross-attention injects the physically indexed knowledge embedding,
as shown in Fig.~\ref{fig:network-architecture-1}.

During sampling, the reverse update is given by
\begin{equation}
\begin{aligned}\boldsymbol{\hat{g}}^{l,(t-1)}_{\mathrm{r}} & =\frac{1}{\sqrt{\alpha_{t}}}(\boldsymbol{\hat{g}}^{l,(t)}_{\mathrm{r}}-\frac{\eta_{t}}{\sqrt{1-\bar{\alpha}_{t}}}f_{\varepsilon}(\boldsymbol{\hat{g}}^{l,(t)}_{\mathrm{r}},\ddot{\mathbf{z}}_{l},t))\\
 & \hspace{0.5cm}+\sqrt{\frac{1-\bar{\alpha}_{t-1}}{1-\bar{\alpha}_{t}}\eta_{t}}\boldsymbol{\varepsilon}_{t}.
\end{aligned}
\label{eq:Forward-Diffusion-Process-1-1-1}
\end{equation}

The stochastic decoder models fine-grained channel variations, while
the map condition preserves the spatial signature associated with
the inferred anchor. It therefore predicts the unobserved beam state
from sparse current observations and the persistent radio-map memory,
which is the predictive function used by the beam-management application.

\subsection{Physics-Informed Training and Memory Update}

The framework does not require accurately labeled UE locations. Instead,
weak and potentially noisy physical cues resolve the global spatial
ambiguity, while spatial clustering, temporal coherence, motion regularization,
and channel reconstruction shape the learned anchors and knowledge
memory.

Let $\mathbf{\bar{p}}^{l}_{\mathrm{r}}$ denote a coarse location
cue obtained from a low-complexity positioning method. The weak alignment
loss is
\begin{equation}
\begin{aligned}\mathcal{L}_{c} & =\sum_{l}||\mathbf{\hat{p}}^{l}_{\mathrm{r}}-\mathbf{\bar{p}}^{l}_{\mathrm{r}}||^{2}.\end{aligned}
\label{eq:Loss1}
\end{equation}
The coarse cues are not treated as ground-truth labels. They only
establish the global coordinate reference required by a physically
indexed service.

To sharpen the spatial assignment over grid cells, we construct an
auxiliary target distribution from assignments $r_{lk}$:
\begin{equation}
\begin{aligned}p_{lk} & =\frac{r^{2}_{lk}/\sum_{l}r_{lk}}{\sum_{k'}r^{2}_{lk'}/\sum_{l}r_{lk'}}.\end{aligned}
\label{eq:Loss2-1}
\end{equation}
The model is trained by minimizing their KL divergence as
\begin{equation}
\begin{aligned}\mathcal{L}_{s} & =\sum_{l}\sum_{k}p_{lk}\mathrm{log}\frac{p_{lk}}{r_{lk}}.\end{aligned}
\label{eq:Loss2}
\end{equation}

Temporal coherence is enforced via a triplet set
\begin{equation}
\mathcal{T}=\{(n,c,f)\in T^{3}:0<|t_{n}-t_{c}|\leq T_{c}<|t_{n}-t_{f}|\}\label{eq:Loss3-1}
\end{equation}
where $t_{n}$, $t_{c}$, and $t_{f}$ denote the timestamps of the
anchor, positive, and negative samples, respectively, and $T_{c}>0$
is a coherence-time threshold that determines temporal closeness.
A triplet loss is thus formulated as
\begin{equation}
\mathcal{L}_{t}=\frac{1}{|\mathcal{T}|}\sum_{(n,c,f)\in\mathcal{T}}\mathrm{max}(||\mathbf{\hat{p}}^{n}_{\mathrm{r}}-\mathbf{\hat{p}}^{c}_{\mathrm{r}}||^{2}-||\mathbf{\hat{p}}^{n}_{\mathrm{r}}-\mathbf{\hat{p}}^{f}_{\mathrm{r}}||^{2}+Q_{t},0)\label{eq:Loss3}
\end{equation}
where $Q_{t}$ is the margin parameter that enforces $\mathbf{\hat{p}}^{n}_{\mathrm{r}}$
to be at least $Q_{t}$ closer to $\mathbf{\hat{p}}^{c}_{\mathrm{r}}$
than to $\mathbf{\hat{p}}^{f}_{\mathrm{r}}$, which preserves local
trajectory neighborhoods in the inferred physical space.

To suppress unrealistic local oscillations, we introduce a physical
dynamics loss based on second-order differences:

\begin{equation}
\begin{aligned}\mathcal{L}_{d} & =\frac{1}{L-2}\sum_{l=2}||\mathbf{\hat{p}}^{l+1}_{\mathrm{r}}-2\mathbf{\hat{p}}^{l}_{\mathrm{r}}+\mathbf{\hat{p}}^{l-1}_{\mathrm{r}}||^{2}.\end{aligned}
\label{eq:Loss4}
\end{equation}
This term promotes smooth motion dynamics without imposing a fixed
mobility model.

Third, reconstruction fidelity is enforced through the diffusion denoising
loss and the radio map commitment loss:
\begin{equation}
\begin{aligned}\mathcal{L}_{r} & =\mathbb{E}_{\boldsymbol{g}_{l},\boldsymbol{\varepsilon},t}[||\boldsymbol{\varepsilon}-f_{\varepsilon}(\boldsymbol{g}^{(t)}_{l},\ddot{\boldsymbol{z}}_{l},t)||^{2}]+||\hat{\boldsymbol{z}}_{l}-\mathrm{sg}(\ddot{\boldsymbol{z}}_{l})||^{2}\end{aligned}
\label{eq:Loss5}
\end{equation}
where $\mathrm{sg}(\cdot)$ is the stop-gradient operator. The first
term trains the diffusion decoder, while the second term keeps the
encoder output consistent with the selected radio-map embedding.

\subsubsection{Model Training}

The encoder and decoder are jointly trained with a composite loss
that integrates all objectives as
\begin{equation}
\begin{aligned}\mathcal{L}_{total} & =\mathcal{L}_{s}+\mathcal{L}_{r}+\lambda_{c}\mathcal{L}_{c}+\lambda_{t}\mathcal{L}_{t}+\lambda_{d}\mathcal{L}_{d}\end{aligned}
\label{eq:Loss5-1}
\end{equation}
where $\lambda_{c}$, $\lambda_{t}$ and $\lambda_{d}$ are weights
to balance each term.

\subsubsection{Radio Map Memory Update}

The memory vectors are updated using an exponential moving average
(EMA) strategy \cite{Razavi:J19}. For each vector $\boldsymbol{e}_{k}$,
let $m_{k}$ be the accumulated sum of the encoder outputs $\hat{\boldsymbol{z}}_{l}$
assigned to $\boldsymbol{e}_{k}$ according to (\ref{eq:Codebook1}),
and $N_{k}$ be the number of times that this vector is selected.
The update rules are given by
\begin{equation}
\begin{aligned}m^{(t)}_{k} & =\gamma m^{(t-1)}_{k}+(1-\gamma)\sum_{i\in\mathcal{I}_{k}}\hat{\boldsymbol{z}}^{(i)}_{l}\\
N^{(t)}_{k} & =\gamma N^{(t-1)}_{k}+(1-\gamma)|\mathcal{I}_{k}|\\
\boldsymbol{e}^{(t)}_{k} & =\frac{m^{(t)}_{k}}{N^{(t)}_{k}+\zeta}
\end{aligned}
\label{eq:Codebook_Update}
\end{equation}
where $\gamma$ is the decay factor, $\mathcal{I}_{k}$ denotes the
set of samples assigned to $\boldsymbol{e}_{k}$, $|\mathcal{I}_{k}|$
is the number of assigned samples, and $\zeta$ is a small constant
to prevent division by zero.

\section{Experiment Results}

We evaluate the proposed model on two datasets: a simulated outdoor
dataset generated by the \ac{rt} software \emph{Wireless InSite}
and a real-world indoor measurement dataset.

\begin{figure}[!t]
\centering\subfigure[]{\label{fig:Env1}\includegraphics[scale=0.24]{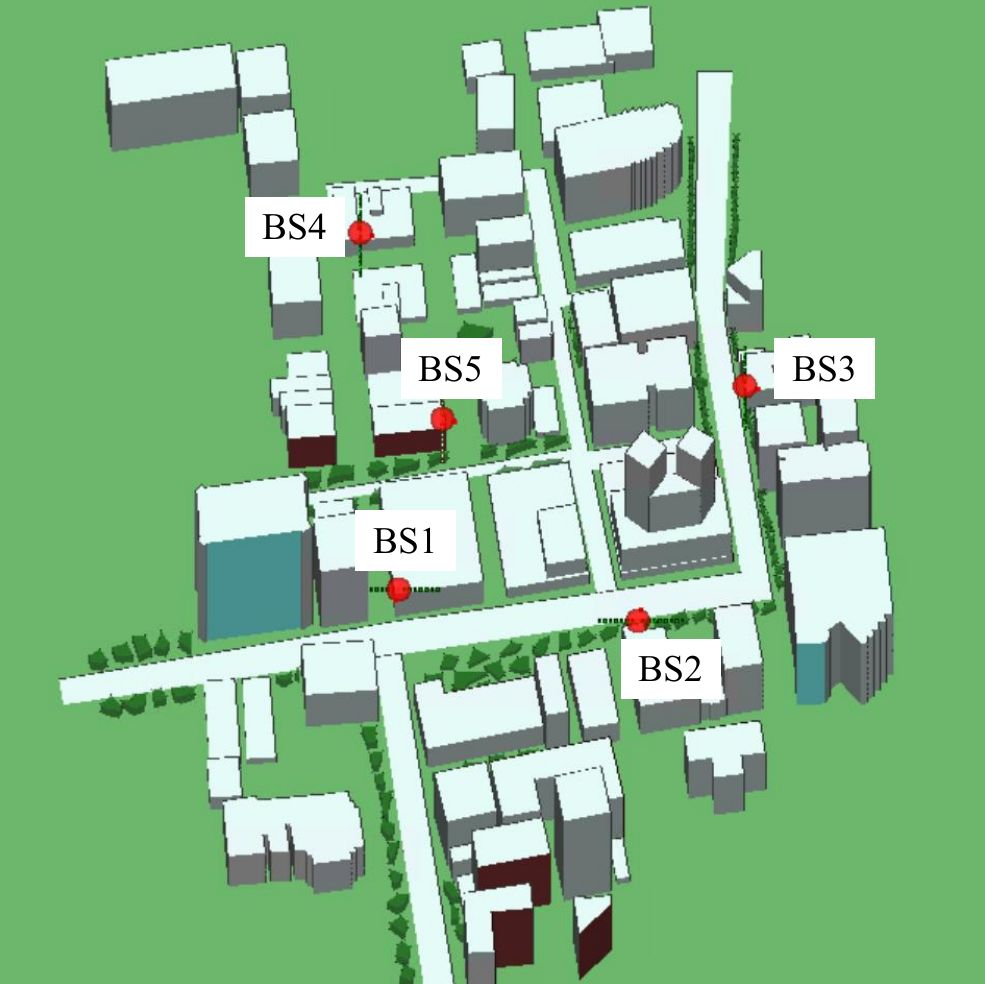}
}\hspace{0.27cm}\subfigure[]{\label{fig:Env2}\includegraphics[scale=0.392]{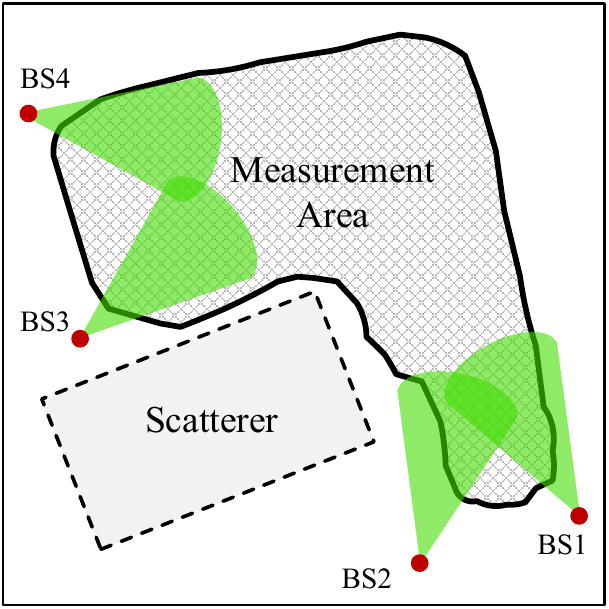}
}\caption{Environmental topology: a) Scenario I: 3D terrain of the outdoor urban
environment; b) Scenario II: top view of the indoor environment \cite{dataset-dichasus-cf0x}.}
\label{fig:Env}
\end{figure}

In the simulated scenario, the urban topology covers a $710$ m $\times$
$740$ m area of San Francisco, USA, as shown in Fig.~\ref{fig:Env1}.
Five \acpl{bs} with 16-element \ac{mimo} arrays are deployed on
rooftops, and a DFT codebook is used. Mobile \acpl{ue} at 2\,m height
follow random-walk trajectories along roads. Channel data are generated
at 2.8 GHz with 100\,MHz bandwidth. In total, \ac{csi} measurements
were collected at 18,844 locations.

In the indoor scenario, we use the DICHASUS dataset~\cite{dataset-dichasus-cf0x}.
The layout is plotted in Fig.~\ref{fig:Env2}. The system includes
four \acpl{tx} with 2\,\texttimes\,4 uniform rectangular arrays.
Channels are measured at 1.272\,GHz with 50\,MHz bandwidth. A mobile
robot with an \ac{rx} at 0.94\,m height traverses an L-shaped area.
In total, 16,778 position-labeled channel samples are recorded.

The following baselines are compared and summarized:
\begin{enumerate}
\item Radio-map-assisted SKF \cite{ZhengChen:J25}: This method uses a preconstructed
radio map as prior information and applies Kalman filtering to infer
trajectories and track CSI.
\item Semi-supervised \Ac{cc} (e.g., \cite{ZhangSaad:C21,KarmanovZan:C21}):
These baselines calibrate the latent chart using a small number of
location labels through the charting loss $\mathcal{L}_{chart}=\lambda_{t}\mathcal{L}_{t}+\lambda_{c}\mathcal{L}_{c}$:
\begin{itemize}
\item \emph{Semi-CC-GT}: Using noise-free ground-truth \ac{rx} locations
in $\mathcal{L}_{c}$ for the ideal supervision scenario;
\item \emph{Semi-CC-Noisy}: Using coarse and noisy \ac{rx} location labels
in $\mathcal{L}_{c}$ to simulate location uncertainty.
\end{itemize}
\item Real-world CC \cite{TanerPal:J25}: It leverages \ac{tx} locations
as weak supervision by associating stronger received signals with
closer \ac{tx}-\ac{rx} proximity, without explicit \ac{rx} labels. 
\item CAM-aided CSI tracking \cite{WuZeng:C23}: It constructs a channel
angle map (CAM), selects $N_{t}/2$ beams by the estimated \ac{aod},
and applies least squares estimation for tracking.
\end{enumerate}

We consider the following metrics: \emph{1) Localization error}: Defined
as $\frac{1}{L}\sum^{L}_{l}||\mathbf{\hat{p}}^{l}_{\mathrm{r}}-\mathbf{p}^{l}_{\mathrm{r}}||$;
\emph{2) 95th percentile} \emph{error}: Defined as the value below
which $95\%$ of the Euclidean distance errors fall; 3) \emph{Trustworthiness
(TW)}: Evaluates whether the local neighborhoods in the latent space
are physically reliable:
\begin{equation}
\begin{aligned}TW(k) & =1-\gamma\sum^{L}_{l}\sum_{j\notin\mathcal{V}_{l},j\in\mathcal{\hat{V}}_{l}}(d(l,j)-k)\end{aligned}
\label{eq:TW}
\end{equation}
where $\mathcal{V}_{l}$ and $\hat{\mathcal{V}}_{l}$ are the sets
of the $k$ nearest neighbors of point $l$ in the real and latent
spaces, $d(l,j)$ is the rank in the real space, and $\gamma=2/(Lk(2L-3k-1))$
is a normalization factor; 4) \emph{Continuity (CT)}: Quantifies how
well real-space neighborhoods are preserved in the latent space:
\begin{equation}
\begin{aligned}CT(k) & =1-\gamma\sum^{L}_{l}\sum_{j\in\mathcal{V}_{l},j\notin\mathcal{\hat{V}}_{l}}(\hat{d}(l,j)-k)\end{aligned}
\label{eq:CT}
\end{equation}
where $\hat{d}(l,j)$ is the rank in the latent space; 5) \emph{Root
mean square error (RMSE) and normalized mean square error (NMSE)}:
Evaluate CSI reconstruction accuracy; 6) \emph{Channel capacity}:
Defined as $\log_{2}(1+P_{\mathrm{t}}|\mathbf{h}^{*}(\tilde{\mathbf{p}})\mathbf{w}|^{2}/\sigma^{2}_{n})$,
where $\mathbf{w}$ is the beamforming vector, $P_{t}$ is the transmit
power, and $\sigma^{2}_{n}$ is the noise variance.

\subsection{Physical Anchoring in Outdoor Environments \label{subsec:Trajectory-Recovery}}

\begin{figure}[!t]
\centering\subfigure[Ground truth]{\includegraphics[scale=0.32]{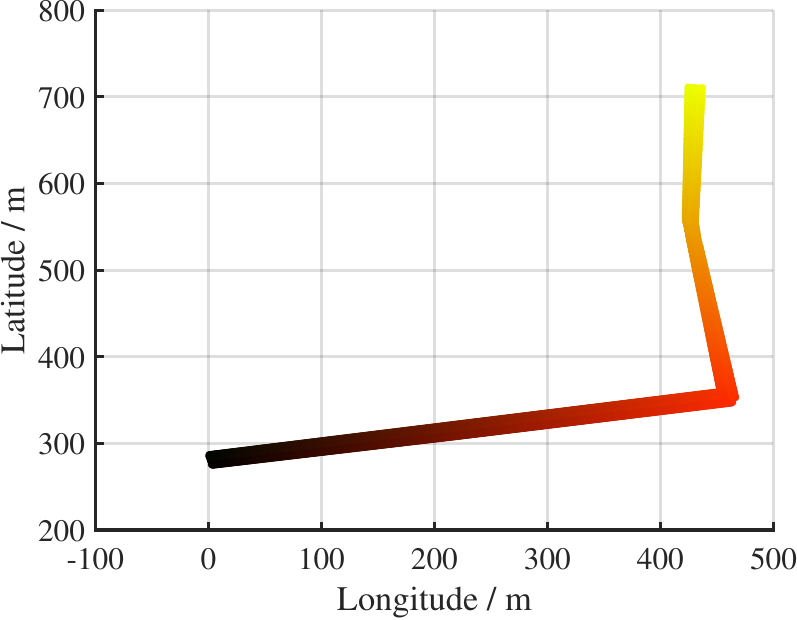}
}\subfigure[The proposed]{\includegraphics[scale=0.32]{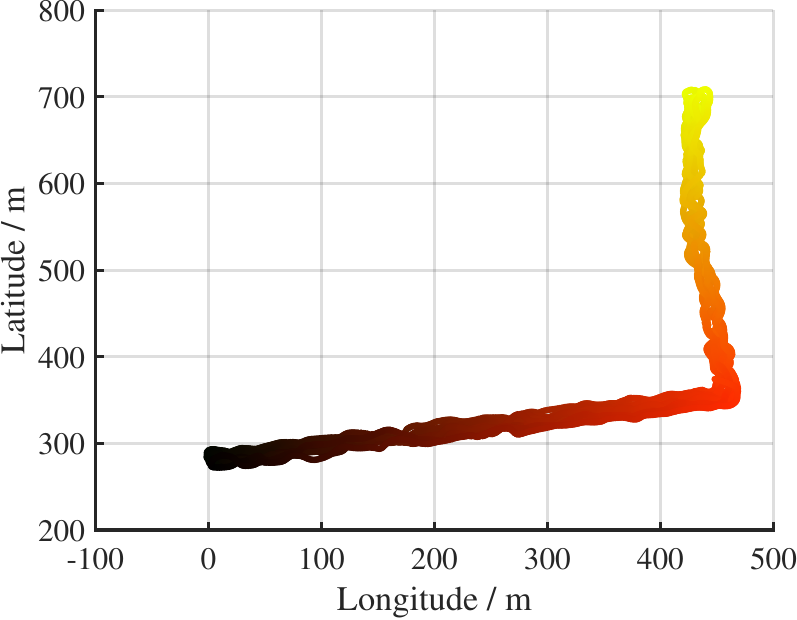}
}

\centering \subfigure[SKF]{\includegraphics[scale=0.32]{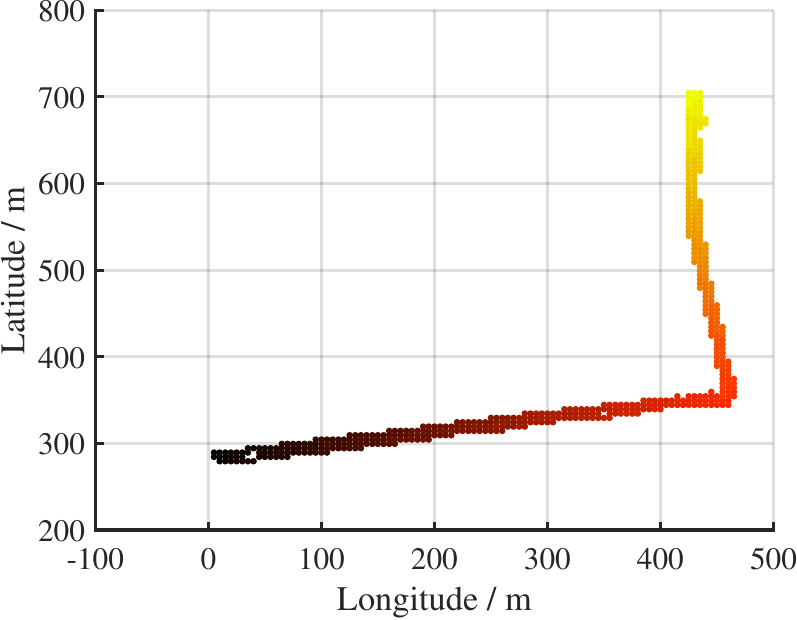}
}\subfigure[Semi-CC-GT]{\includegraphics[scale=0.32]{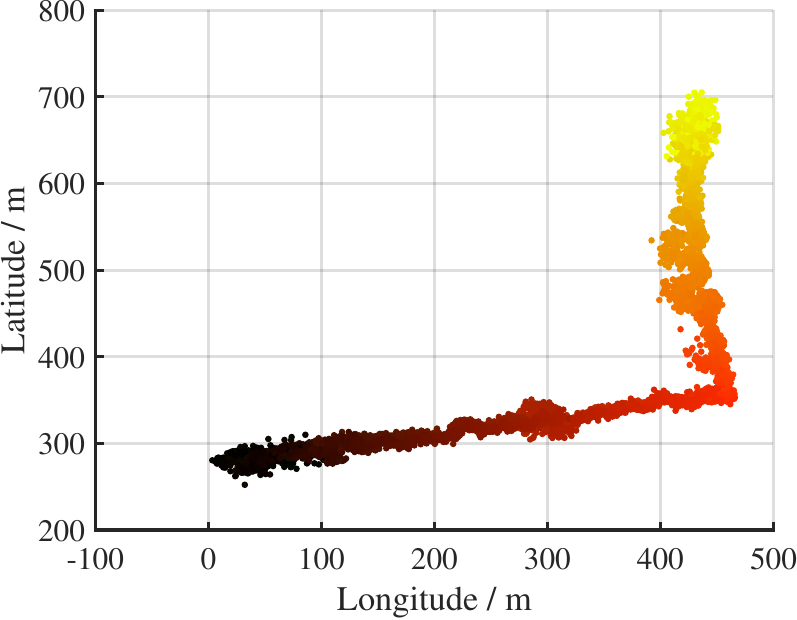}}

\centering \subfigure[Semi-CC-Noisy]{\includegraphics[scale=0.32]{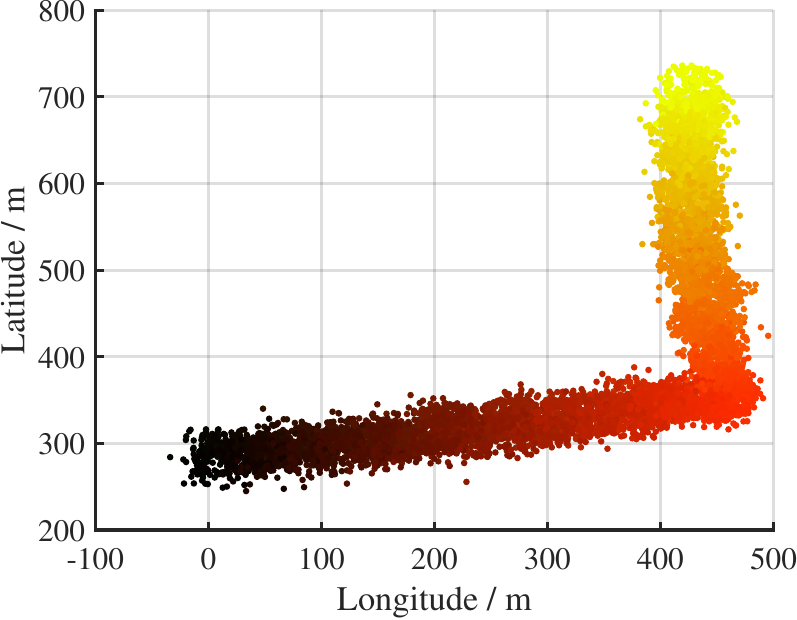}
}\subfigure[Real-world CC]{\includegraphics[scale=0.32]{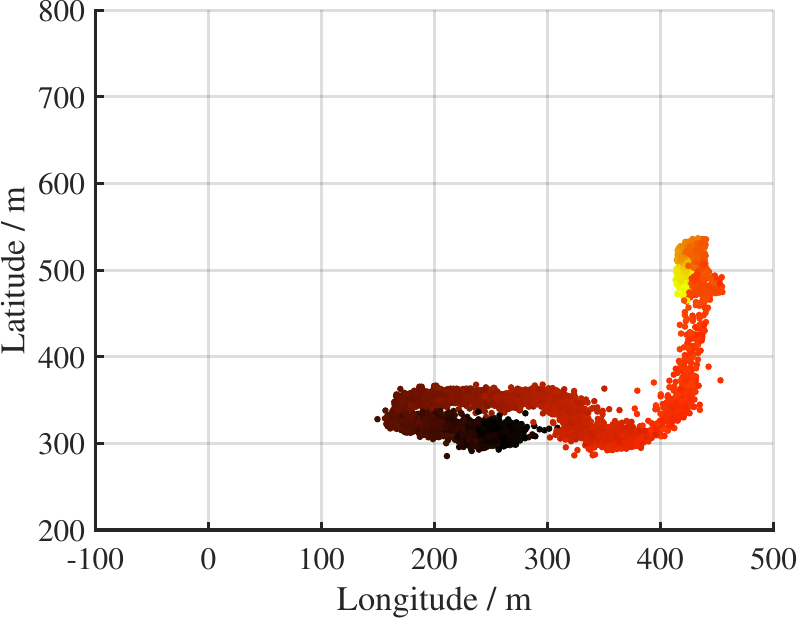}
}\caption{Recovered physical anchors in the outdoor scenario, where the \ac{ue}
moves at 1 m/s along the main road in a back-and-forth pattern. }
\label{fig:Trajectory_Recovery_Outdoor_P1}
\end{figure}

\begin{table}
\caption{Positioning Performance in The Outdoor Environment}

\centering\label{Tab:Trajectory_Outdoor_P1} \renewcommand\arraystretch{1.5}
\begin{tabular}{>{\centering\arraybackslash}m{1.85cm} | >{\centering\arraybackslash}m{1.08cm} | >{\centering\arraybackslash}m{1.08cm} | >{\centering\arraybackslash}m{1.35cm} | >{\centering\arraybackslash}m{1.35cm} }    
\hline    
\multirow{4}{*}{Scheme}  & \multicolumn{2}{c|}{Latent Space Quality}  & \multicolumn{2}{c}{Positioning Error / m} \\   \cline{2-5} & TW~$\uparrow$ & CT~$\uparrow$ & Mean~$\downarrow$ 
& \makecell[c]{\begin{tabular}{@{}c@{\hspace{5pt}}c@{}} \makecell{95th \\ Percentile} & $\downarrow$ \\   \end{tabular} }  \\ 
\hline    SKF                 & 0.986   & 0.987     & 7.24   & 14.78 \\    
\hline    Semi-CC-GT          & 0.990   & 0.994     & 6.81   & 19.57 \\  
\hline    Semi-CC-Noisy       & 0.978   & 0.982     & 14.61   & 29.79 \\  
\hline    Real-world CC       & 0.963   & 0.965     & 82.75   & 217.41 \\ 
\hline    Proposed            & \textbf{0.995} & \textbf{0.995} & \textbf{4.90} & \textbf{9.92} \\ 
\hline 
\end{tabular} 
\end{table}

\begin{figure}[!t]
\centering\subfigure[Ground truth]{\includegraphics[scale=0.32]{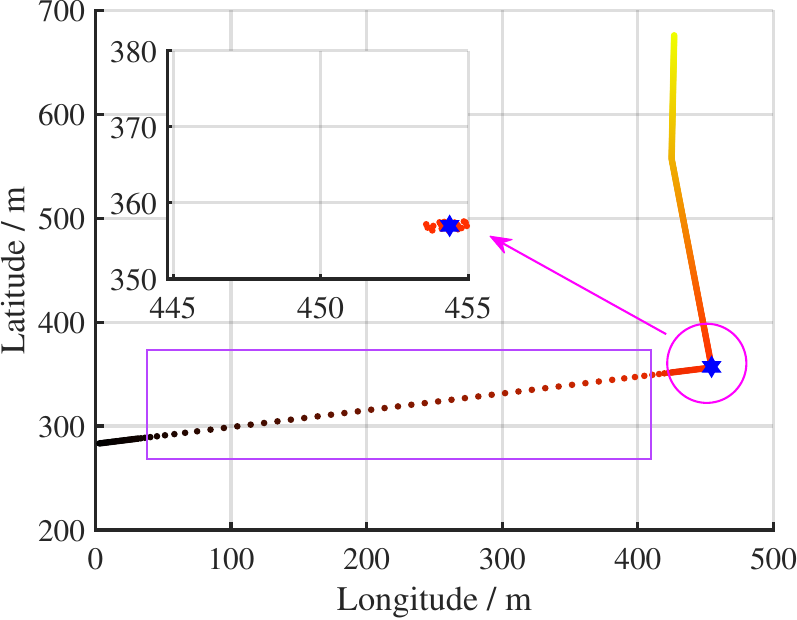}
}\subfigure[The proposed]{\includegraphics[scale=0.32]{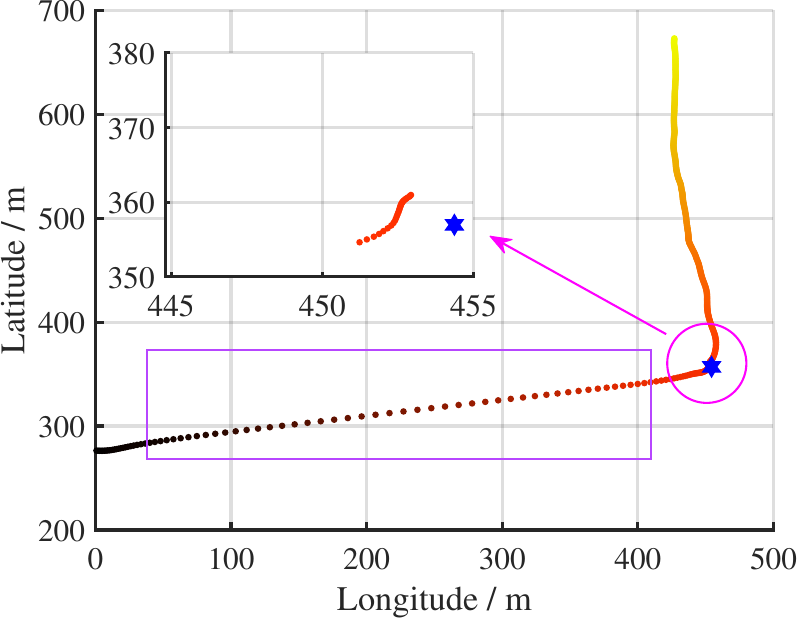}
}

\centering \subfigure[SKF]{\includegraphics[scale=0.32]{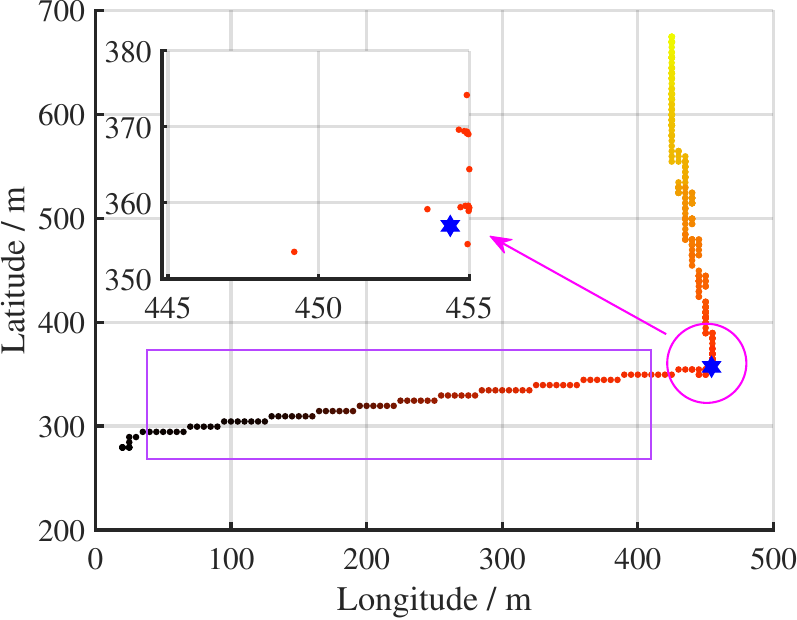}
}\subfigure[Semi-CC-GT]{\includegraphics[scale=0.32]{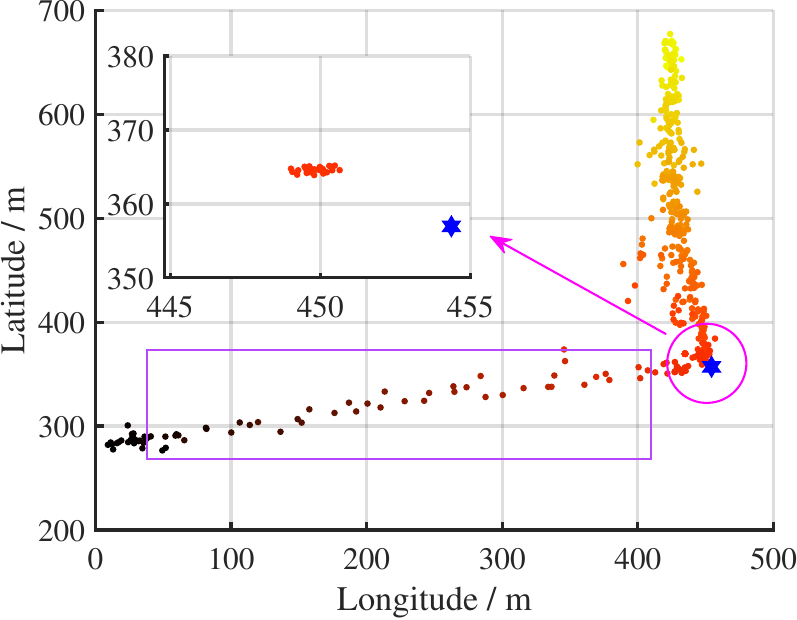}}

\caption{Physical anchor recovery under more complex motion dynamics. The blue
star marks the stop location, while the purple segment indicates acceleration
from 1 m/s to 10 m/s, followed by cruising and subsequent deceleration
back to 1 m/s.}
\label{fig:Trajectory_Recovery_Outdoor_Complex}
\end{figure}

\begin{figure}[!t]
\centering\includegraphics[scale=0.53]{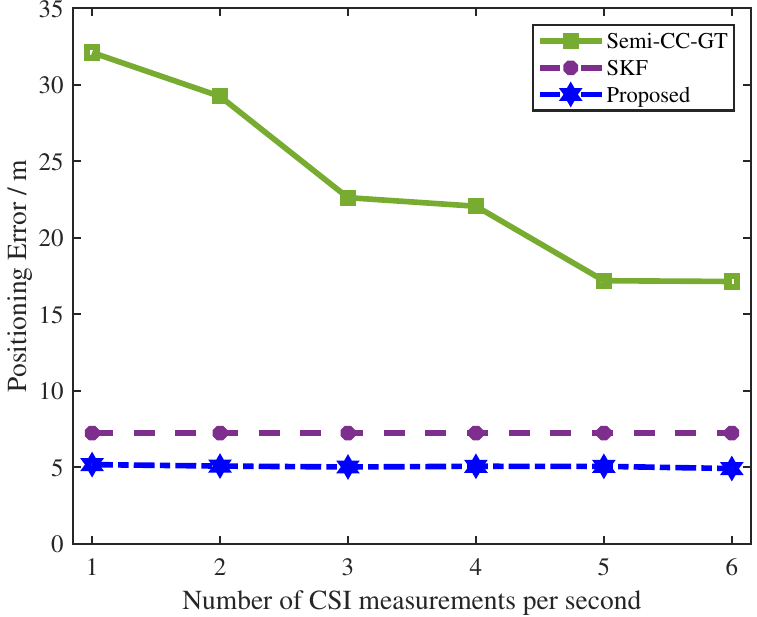} \caption{Positioning error versus the number of measurements per second.}
\label{fig:Sparse_VS_pilot}
\end{figure}

This experiment evaluates whether the proposed framework can recover
reliable intermediate physical anchors from sparse observations. The
grid resolution is set to 5 m. The coarse trajectory $\{\mathbf{\bar{p}}^{l}_{\mathrm{r}}\}^{L}_{l=1}$
is obtained by adding Gaussian noise with variance $\sigma^{2}=400$
to the ground-truth positions. We set $\lambda_{c}=0.01$, $\lambda_{t}=1$,
$\lambda_{d}=50$, $\gamma=0.99$, and the triplet-related parameters
$T_{c}=5$ and $Q_{t}=1$. Semi-CC-GT uses 500 ground-truth location
labels with $\lambda_{c}=1$. Unless otherwise specified, all multi-scale
convolutional layers use kernel sizes of 3, 5, and 7. The mask-aware
embedding dimension, recurrent hidden dimension, and latent dimension
are set to $d_{e}=64$, $d_{r}=128$, and $D=128$, respectively,
and the truncation $\epsilon$ is set to $20\%$ of the trajectory
length.

First, we consider a trajectory moving uniformly along the main road
at 1\,m/s, with \ac{csi} sampled from all \acpl{bs} every 0.1 s.
Fig.~\ref{fig:Trajectory_Recovery_Outdoor_P1} shows the intermediate
physical-anchor recovery results across multiple BSs. It is observed
that the proposed method produces a continuous trajectory with better
temporal and spatial coherence, while SKF yields only discrete points
along the path. Semi-CC-GT benefits from ground-truth location supervision
but still exhibits local disturbances, whereas Semi-CC-Noisy suffers
from global deviations due to noisy calibration labels. Real-world
CC performs the worst. The quantitative comparison is presented in
Table~\ref{Tab:Trajectory_Outdoor_P1}. The proposed method outperforms
all baselines across all metrics, reducing localization error by $32\%$
over SKF and $28\%$ over Semi-CC-GT. Semi-CC-GT performs comparably
to the proposed method in TW and CT, with differences below 0.004.
Although Real-world CC yields a large localization error, its TW and
CT remain relatively high, indicating that the latent spatial structure
is still well preserved.

Second, we evaluate the robustness of the temporal inference mechanism
under a complex trajectory, as shown in Fig.~\ref{fig:Trajectory_Recovery_Outdoor_Complex}.
The \ac{ue} first moves at 1 m/s, pauses for 30 s, and accelerates
from 1 m/s up to 10 m/s with an acceleration rate of 1 m/s\texttwosuperior .
It maintains the high-speed motion for a period before decelerating
back to 1 m/s. The stop position (blue star) corresponds to repeated
samples at the same point, with small perturbations added for clearer
visualization. The proposed method tracks both acceleration and deceleration
phases more accurately than the baselines and keeps the repeated samples
within the same local region. This demonstrates that the data-driven
temporal encoder can learn complex mobility patterns without relying
on a fixed motion model. By contrast, SKF is less effective under
complex mobility, while Semi-CC-GT exhibits local instability despite
using location supervision.

We further evaluate robustness to sparse measurements by varying the
number of \ac{csi} measurements $M$ per second. Fig.~\ref{fig:Sparse_VS_pilot}
shows the positioning error versus the number of measurement $M$.
SKF shows no performance change as $M$ decreases, while the proposed
method exhibits only a minor degradation of about 0.2 meters even
at $M=1$. This robustness stems from our feature extractor, which
stabilizes channel signatures by leveraging angular dependencies and
inter-sample correlations. The proposed method consistently outperforms
SKF across all settings. In contrast, Semi-CC-GT suffers a significant
drop under sparse measurements and improves as $M$ increases, indicating
strong dependence on dense measurements.

\begin{figure}[!t]
\centering\subfigure[Ground truth]{\includegraphics[scale=0.32]{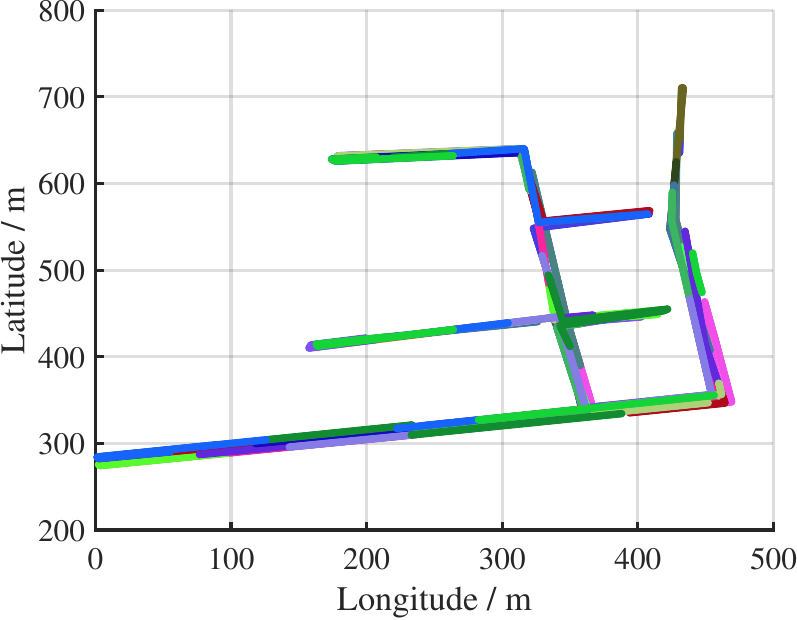}
}\subfigure[The proposed]{\includegraphics[scale=0.32]{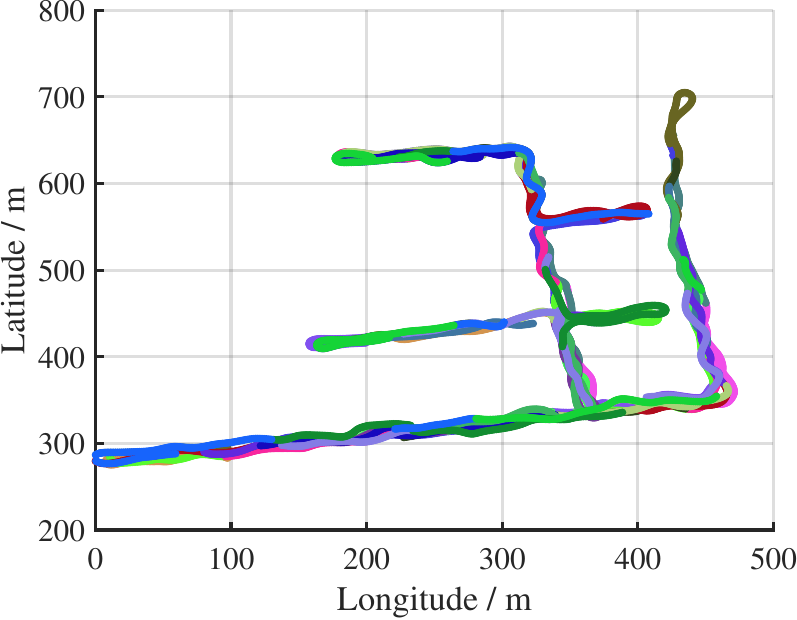}
}

\centering \subfigure[LSTM]{\includegraphics[scale=0.32]{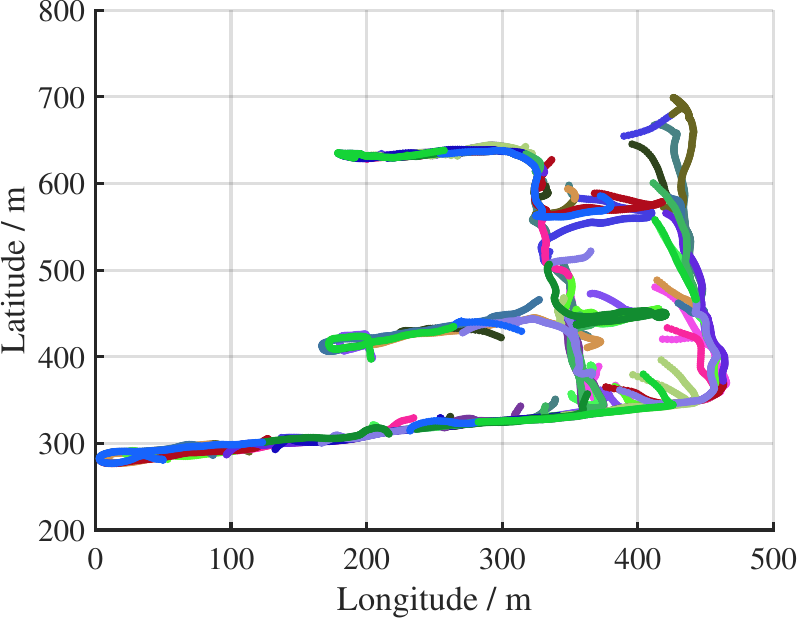}
}\subfigure[Semi-CC-GT]{\includegraphics[scale=0.32]{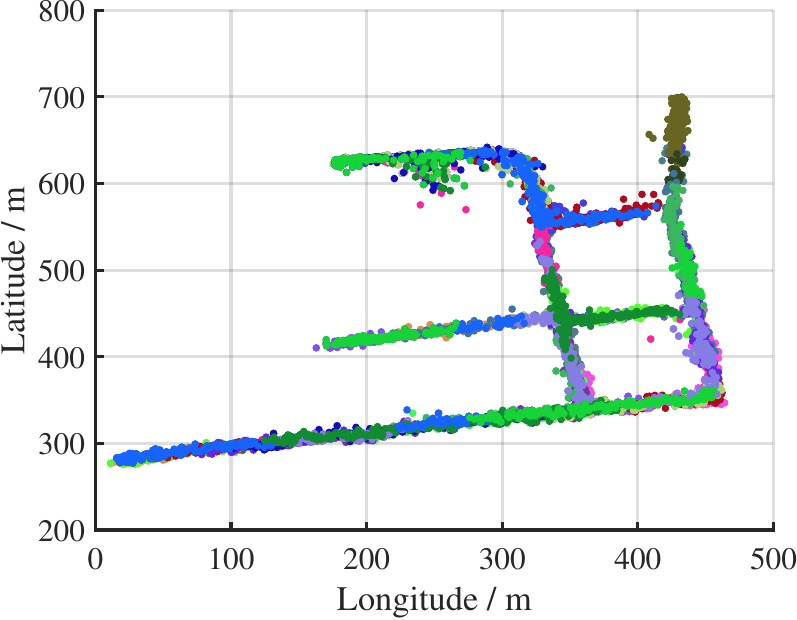}}

\caption{Generalization results across trajectories of varying lengths.}
\label{fig:Discontinuous_Trajectories}
\end{figure}

\subsection{Generalization to Variable-Length Measurement Streams}

This experiment evaluates whether the proposed framework can generalize
to trajectories with varying lengths, because the trajectory length
is usually not fixed and the model should not rely on a complete trajectory
with a predetermined duration. Over 10,000 randomly generated trajectories
with lengths ranging from 200 m to 1,000 m are used for training,
and an additional 100 trajectories are used for evaluation. We include
a conventional long short-term memory (LSTM) baseline.

The generalization results for 40 trajectory samples are shown in
Fig.~\ref{fig:Discontinuous_Trajectories}, and the quantitative
comparison is summarized in Table~\ref{Tab:Trajectory_Outdoor_P2}.
The proposed method consistently outperforms the baselines, achieving
higher TW and CT scores and more than a 2.5 m reduction in mean localization
error. Compared with the LSTM, we improve localization accuracy by
53\% and reduce the 95th-percentile error by 70\%. These results indicate
that the proposed model provides more stable intermediate physical
anchors under variable lengths.

Fig.~\ref{fig:Discontinuous_Trajectories} illustrates the difference
between the proposed design and the LSTM baseline. The trajectories
of the proposed method are smoother, more continuous, and more closely
aligned with the ground truth. In contrast, Semi-CC-GT shows local
fluctuations and LSTM exhibits clear deviations near the beginning
of each trajectory. We further analyze the length of the initial segments
of LSTM where the positioning error exceeded the threshold of the
mean plus one standard deviation. Our statistics revealed that these
highly inaccurate initial predictions account for approximately 13.4\%
of the total trajectory length on average. We conservatively chose
20\% to ensure the removal of the most unreliable initial states.

\begin{table}
\caption{Generalization Results for Trajectories of Arbitrary Lengths}

\centering\label{Tab:Trajectory_Outdoor_P2} \renewcommand\arraystretch{1.5}
\begin{tabular}{>{\centering\arraybackslash}m{1.85cm} | >{\centering\arraybackslash}m{1.08cm} | >{\centering\arraybackslash}m{1.08cm} | >{\centering\arraybackslash}m{1.33cm} | >{\centering\arraybackslash}m{1.35cm} }    
\hline    
\multirow{4}{*}{Scheme}  & \multicolumn{2}{c|}{Latent Space Quality}  & \multicolumn{2}{c}{Positioning Error / m} \\   \cline{2-5} & TW~$\uparrow$ & CT~$\uparrow$ & Mean~$\downarrow$ 
& \makecell[c]{\begin{tabular}{@{}c@{\hspace{5pt}}c@{}} \makecell{95th \\ Percentile} & $\downarrow$ \\   \end{tabular} }  \\ 
\hline    Semi-CC-GT          & 0.992   & 0.992     & 7.64   & 19.61 \\  
\hline    LSTM                & 0.991   & 0.989     & 10.27   & 28.77 \\  
\hline    Proposed            & \textbf{0.995} & \textbf{0.995} & \textbf{4.81} & \textbf{8.59} \\ 
\hline 
\end{tabular} 
\end{table}

\subsection{Real-Measurement Validation}

\begin{figure}[!t]
\centering\subfigure[Ground truth]{\includegraphics[scale=0.32]{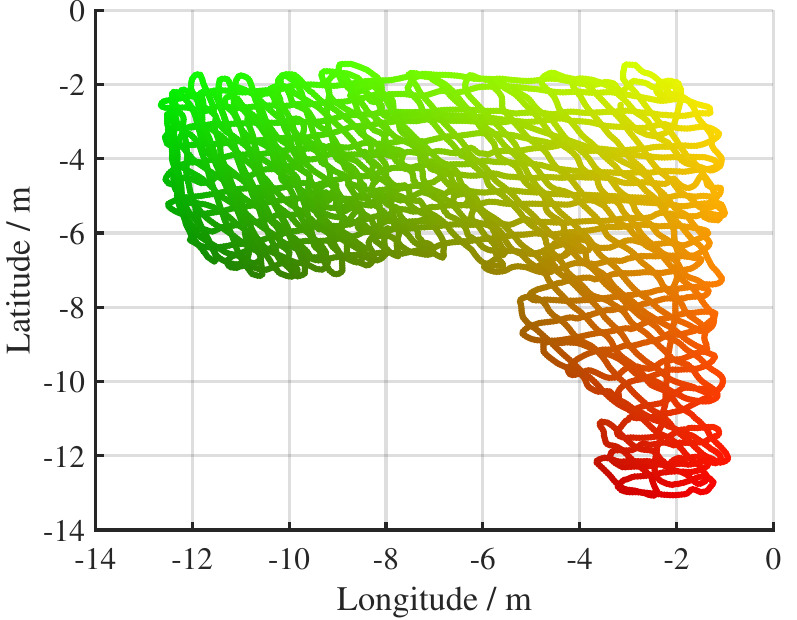}
}\subfigure[Imprecise trajectory]{\includegraphics[scale=0.32]{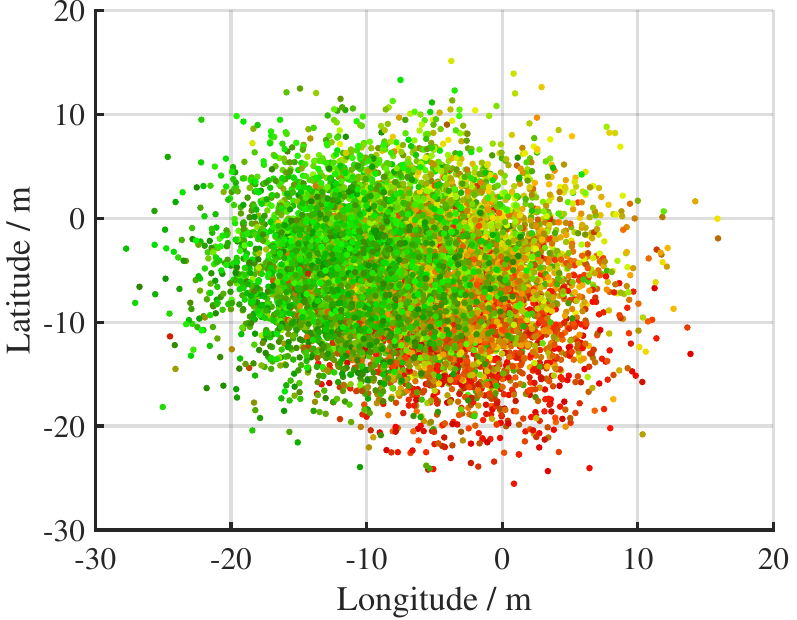}
}

\centering \subfigure[The proposed]{\includegraphics[scale=0.32]{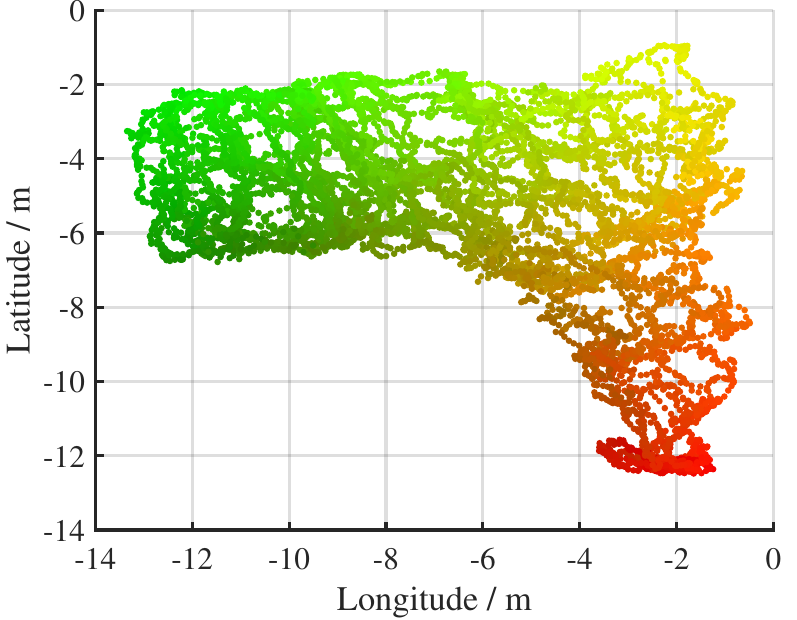}
}\subfigure[SKF]{\includegraphics[scale=0.32]{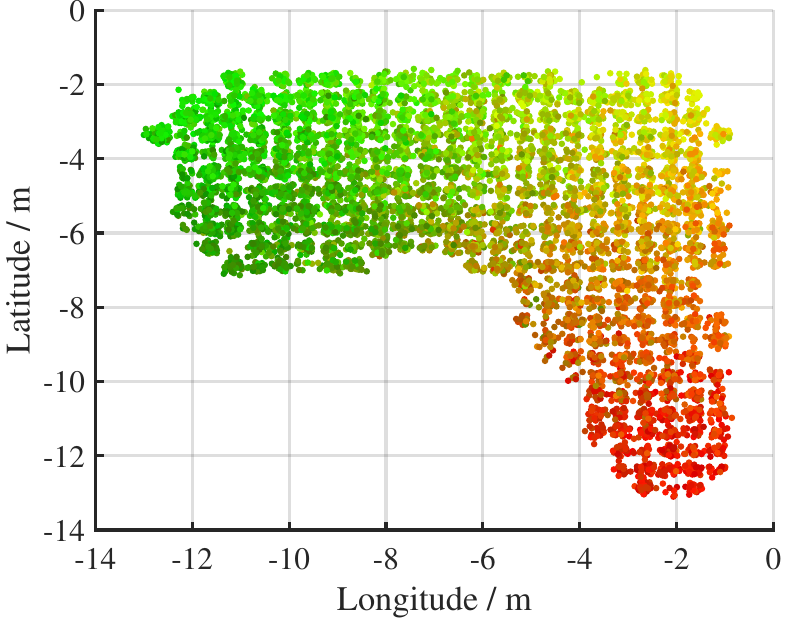}}

\centering \subfigure[Semi-CC-Noisy]{\includegraphics[scale=0.32]{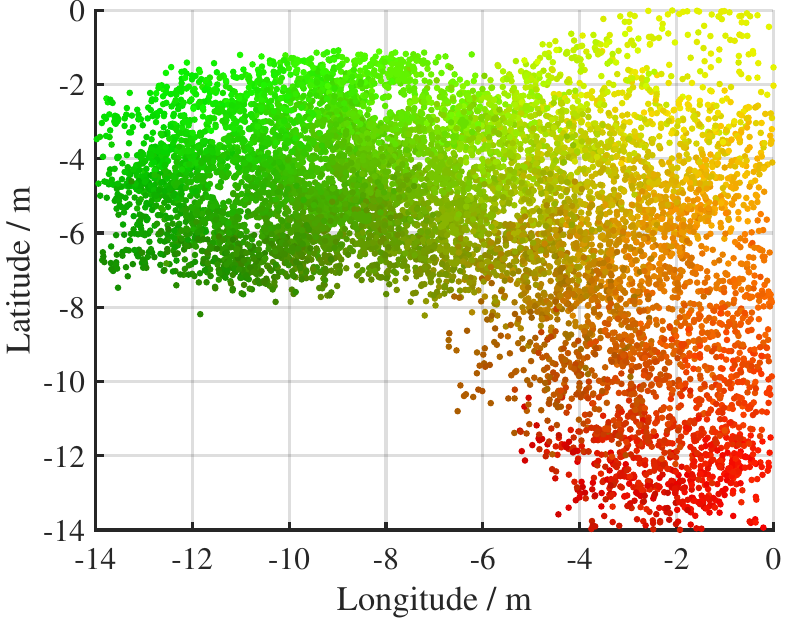}
}\subfigure[Real-world CC]{\includegraphics[scale=0.32]{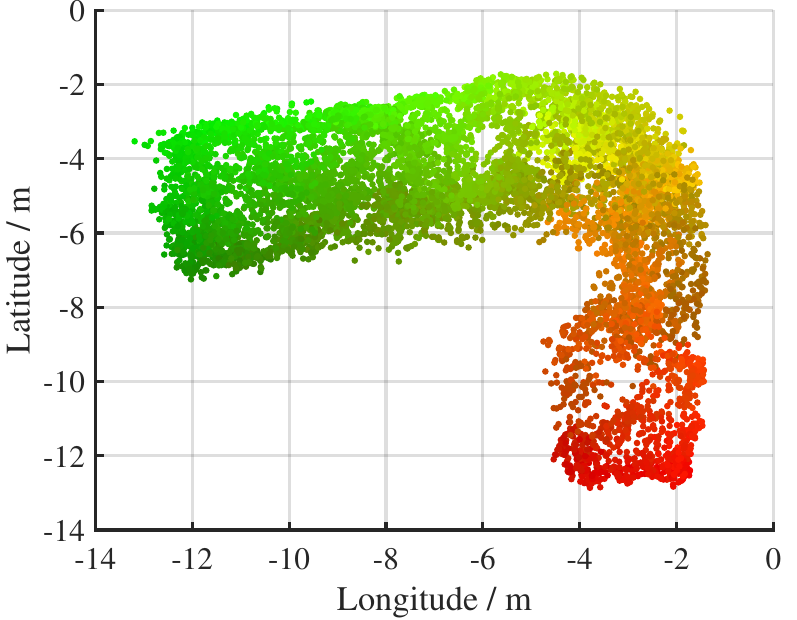}
}\caption{Physical-anchor recovery using real indoor channel measurements.}
\label{fig:Trajectory_Recovery_Indoor}
\end{figure}

In this scenario, the radio map grid resolution is set to 0.5 m, and
the loss weights are configured as $\lambda_{c}=0.01$, $\lambda_{t}=5$,
and $\lambda_{d}=2$. The imprecise trajectory $\{\mathbf{\bar{p}}^{l}_{\mathrm{r}}\}^{L}_{l=1}$
is generated by adding Gaussian noise with variance $\sigma^{2}=25$. 

Fig.~\ref{fig:Trajectory_Recovery_Indoor} compares the recovered
trajectories of different methods. The initial trajectory is heavily
corrupted by noise, which obscures positional information and makes
direct localization challenging. Although all methods can roughly
separate samples along the green-to-red gradient region, only the
proposed method maintains trajectory continuity and smoothness, closely
matching the ground truth, whereas the baselines produce scattered
clusters with weaker temporal coherence.

\begin{table}
\caption{Positioning Performance Comparison for Indoor Scenario}

\centering\label{Tab:Trajectory_Indoor} \renewcommand\arraystretch{1.5}
\begin{tabular}{>{\centering\arraybackslash}m{1.85cm} | >{\centering\arraybackslash}m{1.08cm} | >{\centering\arraybackslash}m{1.08cm} | >{\centering\arraybackslash}m{1.35cm} | >{\centering\arraybackslash}m{1.35cm} }    
\hline    
\multirow{4}{*}{Scheme}  & \multicolumn{2}{c|}{Latent Space Quality}  & \multicolumn{2}{c}{Positioning Error / m} \\   \cline{2-5} & TW~$\uparrow$ & CT~$\uparrow$ & Mean~$\downarrow$ 
& \makecell[c]{\begin{tabular}{@{}c@{\hspace{5pt}}c@{}} \makecell{95th \\ Percentile} & $\downarrow$ \\   \end{tabular} }  \\ 
\hline    SKF                 & 0.847  & 0.886    & 1.56   & 3.19 \\    
\hline    Semi-CC-Noisy       & 0.975   & 0.983     & 0.76   & 1.63 \\  
\hline    Real-world CC       & 0.969  & 0.988      & 0.99   & 2.28 \\ 
\hline    Proposed            & \textbf{0.991} & \textbf{0.992} & \textbf{0.49} & \textbf{1.08} \\ 
\hline 
\end{tabular} 
\end{table}

\begin{figure}[!t]
\centering\hspace{0.1cm}\includegraphics[scale=0.53]{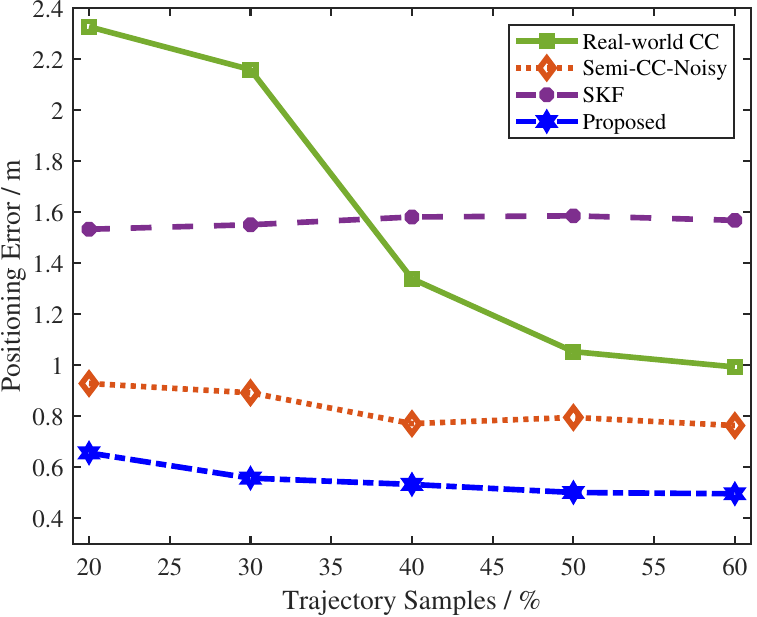}
\caption{Positioning error versus the ratio of available trajectory samples.}
\label{fig:SampleVsLength}
\end{figure}

Table~\ref{Tab:Trajectory_Indoor} summarizes the localization accuracy
in the indoor scenario. The proposed model consistently outperforms
all baselines across all metrics, with TW and CT scores improving
by at least 0.01. It reduces localization error by $35.1\%$ to $68.3\%$
and the 95th-percentile error by $32.9\%$ to $66.1\%$. Notably,
Real-world CC outperforms SKF here, likely due to slow channel variations
and dominant LOS conditions indoors.

We further evaluate positioning error across various trajectory sampling
ratios from 20\% to 60\% of the indoor dataset. As shown in Fig.~\ref{fig:SampleVsLength},
the positioning errors decrease for all methods except SKF as the
number of samples increases. Real-world CC improves significantly
when the sample ratio exceeds $50\%$ but performs worse than SKF
below $40\%$. In contrast, the proposed method maintains high localization
accuracy even with limited samples, demonstrating robustness and efficiency
in data-scarce scenarios and thus reducing the practical data collection
burden.

\subsection{Physically Indexed MIMO Beam Map Construction}

We evaluate the performance of the proposed model in constructing
MIMO beam maps from sparse CSI measurements. 

\begin{figure}[!t]
\centering\subfigure[Ground truth]{\includegraphics[scale=0.345]{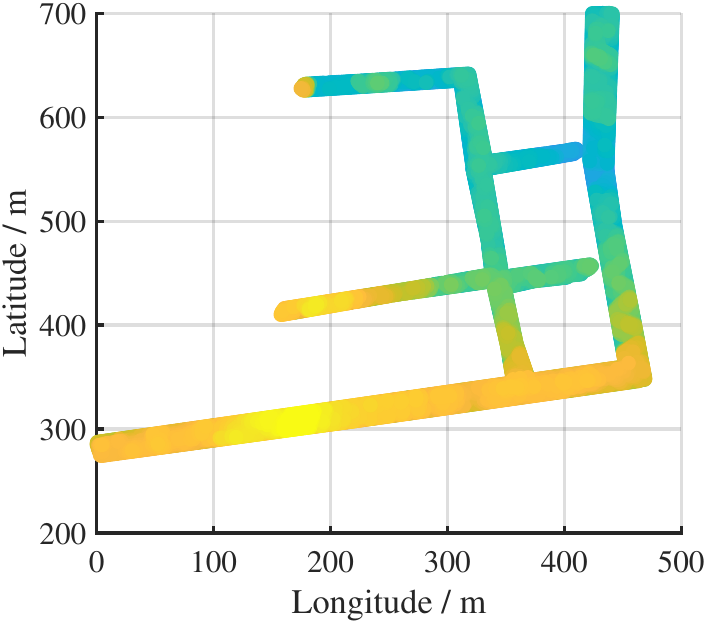}}\subfigure[The proposed]{\includegraphics[scale=0.345]{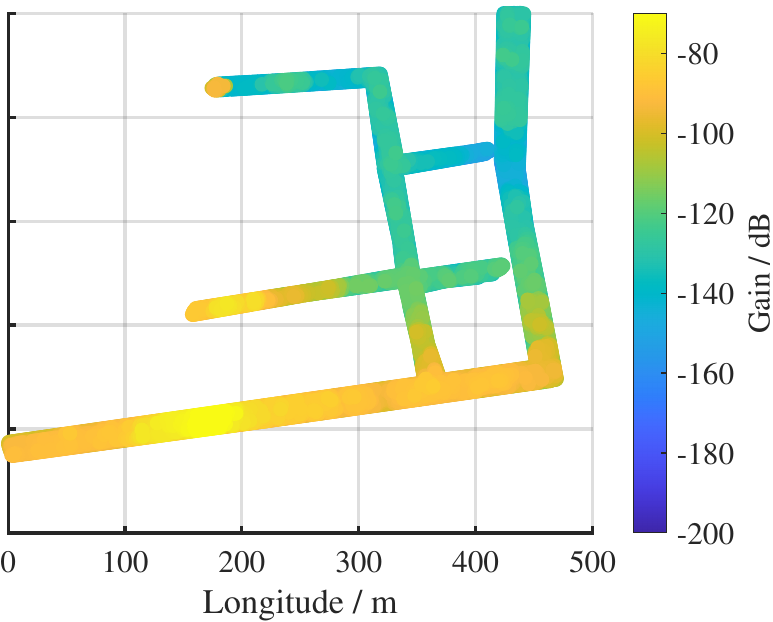}}

\centering\subfigure[Ground truth]{\includegraphics[scale=0.345]{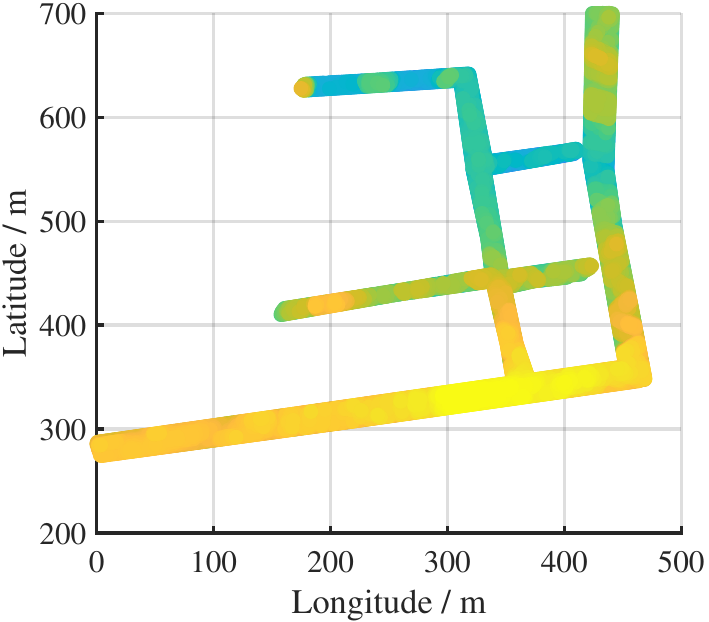}}\subfigure[The proposed]{\includegraphics[scale=0.345]{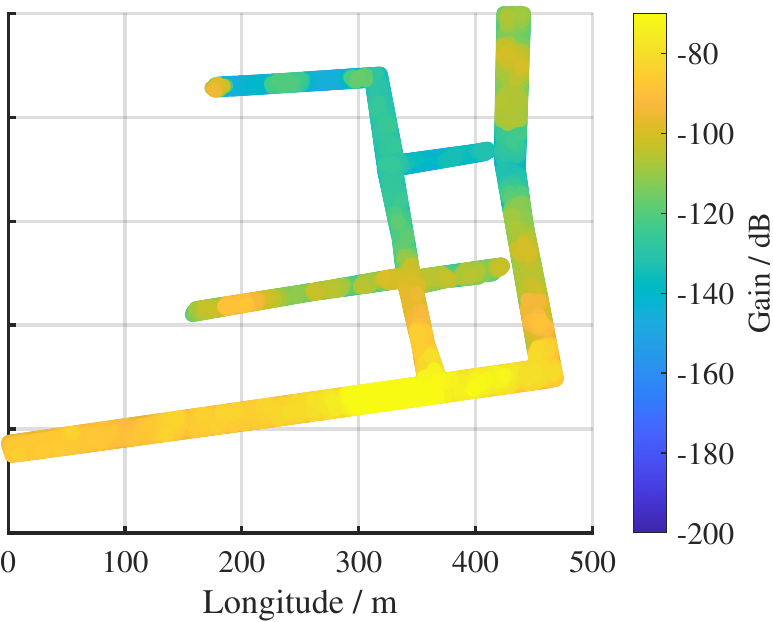}}\caption{MIMO beam map of BS1 for specific beam directions: a) and c) The ground
truth; b) and d) The beam map generated by the proposed model.}
\label{fig:RadioMap}
\end{figure}

Fig.~\ref{fig:RadioMap} shows the reconstructed MIMO beam maps of
BS1 for two beam directions along the main road. In dense urban environments,
radio propagation exhibits complex spatial patterns due to blockage,
reflection, and multipath effects. The results show that the proposed
model can reconstruct the main beam geometry, including beam shape,
direction, blockage and reflections. The close agreement with the
ground truth indicates that the learned radio-map embedding provides
effective spatial signatures for high-fidelity CSI generation. 

Table~\ref{Tab:RadioMapConstruction} summarizes the radio map reconstruction
performance in terms of NMSE and RMSE under varying numbers of \ac{csi}
measurements (i.e., $M=$ 1, 3, and 5). As $M$ increases, the accuracy
of the proposed method improves, while SKF remains largely unchanged.
Notably, even with $M=1$, the proposed method outperforms SKF, reducing
NMSE by 52.3\% and RMSE by 26.5\%. With $M=5$, the reductions further
increase to 59.1\% and 34.3\%, respectively. These results demonstrate
that the proposed method can reconstruct MIMO beam maps from extremely
sparse measurements, even with only one CSI measurement per BS at
each UE location, showing its robustness under limited channel observations.

\begin{table}
\caption{Comparison Results of Radio Map Construction.}
\centering\label{Tab:RadioMapConstruction}\renewcommand\arraystretch{1.5}
\begin{tabular}{ p{1.5cm}<{\centering}| p{1.85cm}<{\centering}| p{1.85cm}<{\centering}| p{1.85cm}<{\centering}}   
\hline   

\multirow{3}{*}{Scheme}  &\multicolumn{3}{c}{The number of CSI measurements}  \\ 

\cline{2-4}
                   & M=1      & M=3    & M=5   \\  
\cline{2-4}
                   &\multicolumn{3}{c}{NMSE / RMSE}   \\  
\hline   

SKF                & 0.013 / 0.068            & 0.012 / 0.067    & 0.012 / 0.067   \\    
\hline   

Proposed           & \textbf{0.0062 / 0.050}   & \textbf{0.0056 / 0.047}    & \textbf{0.0049 / 0.044}\\    
\hline  

\end{tabular} 
\end{table}

\begin{figure*}[!t]
\centering\subfigure[]{\includegraphics[scale=0.4]{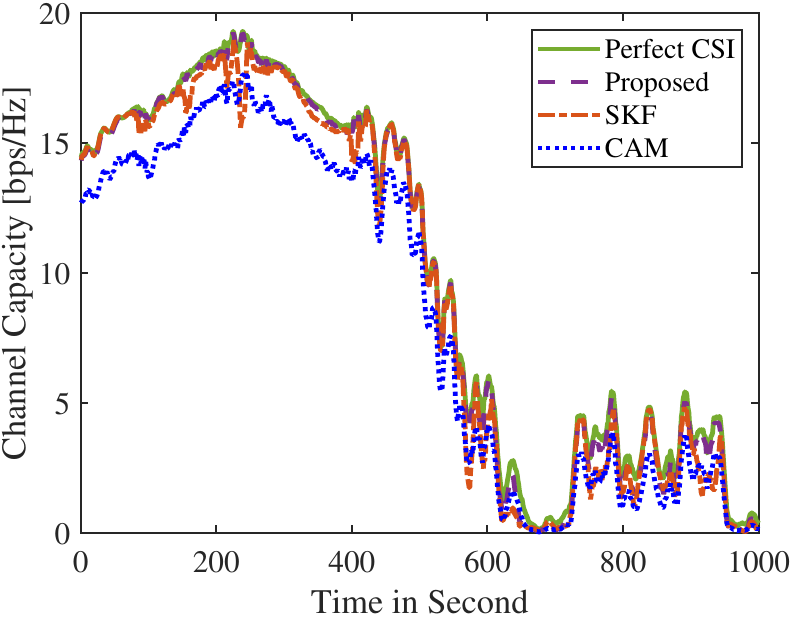}}\hspace{0.5cm}\subfigure[]{\includegraphics[scale=0.4]{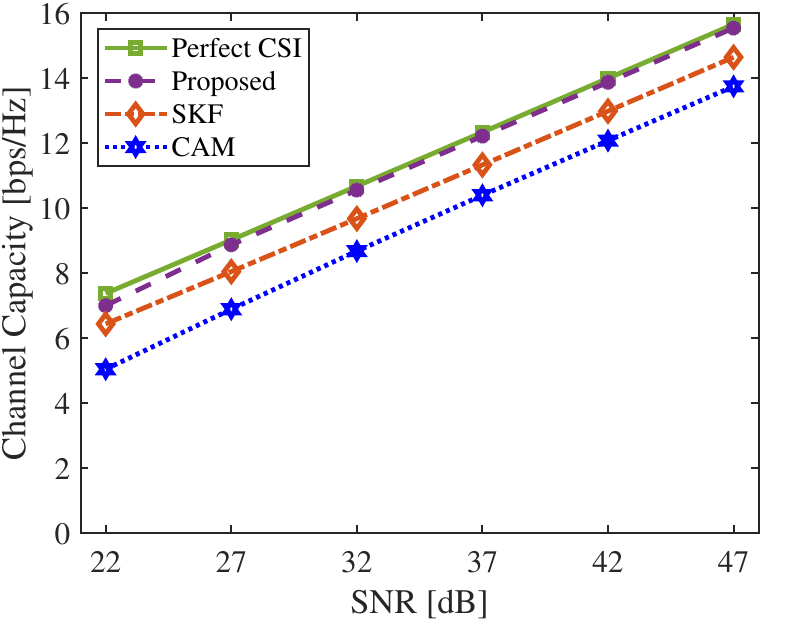}}\hspace{0.6cm}\subfigure[]{\includegraphics[scale=0.4]{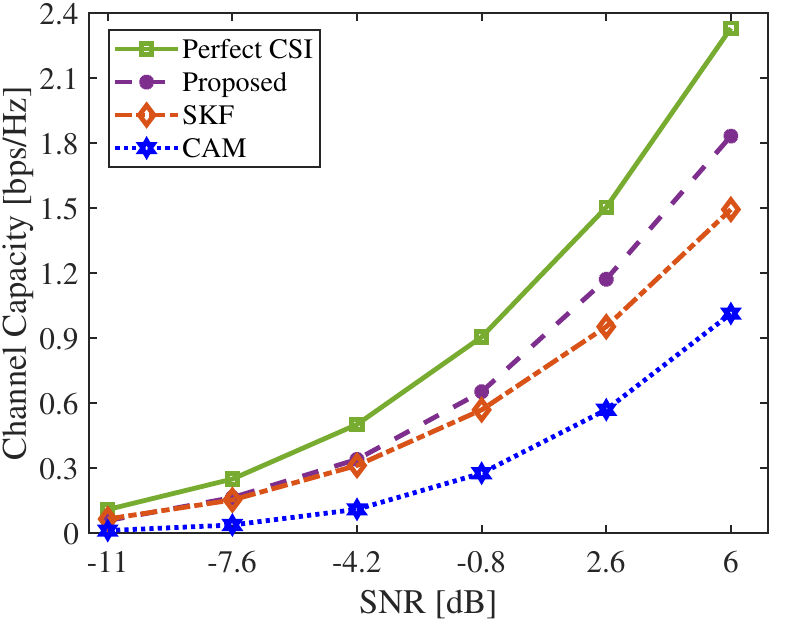}}\caption{Radio-map-embedded beam tracking under LOS and NLOS propagation. a)
Channel capacity for a trajectory transitioning through LOS and NLOS
propagation conditions to BS1; b) Channel capacity for LOS trajectories
to BS2; c) Channel capacity for NLOS trajectories to BS3.}
\label{fig:CSI_tracking}
\end{figure*}

\subsection{Channel-Knowledge-Assisted Beam Tracking}

This section shows the real-time application of the proposed method
for radio-map-embedded beam tracking.

We consider different SNR levels, where the SNR is defined as $\mathrm{SNR}=P_{t}\mathbb{E}[|\mathbf{h}^{*}(\tilde{\mathbf{p}})\mathbf{w}|^{2}]/\sigma^{2}_{n}$
based on the mean channel power and the given noise variance. The
outdoor dataset is used for beam tracking evaluation, and the parameters
are kept consistent with Section~\ref{subsec:Trajectory-Recovery}.

Fig.~\ref{fig:CSI_tracking}(a) shows the real-time channel capacity
along a trajectory from LOS to NLOS regions. The channel exhibits
clear temporal fluctuations, and all methods generally capture the
capacity variations with trends close to the perfect-CSI benchmark.
Among them, the proposed method most closely matches the perfect CSI
with the minimal deviations, outperforming SKF and CAM in both LOS
and NLOS regions. 

Fig.~\ref{fig:CSI_tracking}(b) and (c) illustrate the average channel
capacity along trajectories under LOS and NLOS conditions, respectively.
While all methods show improved channel capacity with increasing SNR,
the proposed method outperforms the baselines, and achieves performance
closest to the perfect CSI, reaching up to $98\%$ of the ideal channel
capacity in LOS conditions and maintaining approximately $77.5\%$
even in NLOS conditions. By contrast, SKF reaches $97\%$ of the perfect
CSI channel capacity in LOS conditions, but drops to only $57\%$
in NLOS regions. The advantage of the proposed method in NLOS condition
becomes more pronounced at higher SNRs, improving channel capacity
over SKF by $25.1\%$ at SNR $=-11$ dB and up to $35.4\%$ at SNR
$=6$ dB.

\subsection{Complexity Analysis}

The proposed framework contains 5.21 M trainable parameters. For a
sequence with length $L$, the encoder complexity scales linearly
with $L$, i.e., $\mathcal{O}(L)$. The radio-map retrieval has an
upper-bound complexity of $\mathcal{O}((K+D)L)$. The diffusion decoder
requires complexity $\mathcal{O}(TL)$ for generating a full sequence
with $T$ denoising steps. By contrast, SKF requires maintaining and
updating a high-dimensional state covariance matrix, resulting in
a higher computational complexity of approximately $\mathcal{O}(D^{3}_{s}L)$,
where $D_{s}$ is the state dimension.

The runtime is evaluated on a PC with an Intel Core i7-8700K CPU @
3.70 GHz, 32 GB RAM, and an NVIDIA Quadro P4000 GPU. In sliding-window
inference with $L=500$, the encoder takes 36.93 ms on average for
trajectory recovery, while the diffusion decoder takes 738.50 ms for
one CSI sample generation with $T=1000$. In comparison, SKF requires
approximately 30 s to track the same sequence. These results show
that the proposed framework provides much faster CSI tracking than
SKF, while the main computational bottleneck lies in the diffusion-based
CSI generation stage.

\section{Conclusion}

This paper developed a self-localizing, radio-map-embedded generative
framework that organizes highly sparse CSI measurements into a hierarchical
wireless memory without requiring explicit location labels. The proposed
framework addressed three key challenges. First, to avoid dependence
on location labels, intermediate trajectory inference is used as a
physical anchoring mechanism that aligns sparse CSI observations with
a physically indexed radio map memory, coupling self-localization
and long-term channel knowledge learning. Second, to reduce acquisition
and processing overhead, beam-domain RSS was theoretically established
as a compact spatial signature, while a dual-scale feature extractor
was designed to recover robust angular and inter-sample representations
from sparse observations. Third, to address complex mobility and error
accumulation, a hybrid temporal encoder further consolidates recent
CSI into stable short-term context to reduce boundary bias and suppress
channel-induced jitters. The inferred anchors spatially index and
update the long-term radio-map embedding, whose retrieved channel
representation conditions a diffusion decoder for memory-assisted
CSI reconstruction and downstream beam tracking. Experimental results
demonstrate that the proposed model can improve localization accuracy
by over $30\%$ and achieve a $20\%$ channel capacity gain in NLOS
scenarios compared to Kalman filter methods.

\begin{appendices}

\section{Proof of Theorem 1}\label{app:Theom proof}

For the $j$th beam, the high-SNR dominant-path approximation gives
$\rho(\tilde{\mathbf{p}},\mathbf{w}_{j})\approx\mu_{j}+n_{j}$, where
$\mu_{j}=\beta_{D}\mathbf{a}^{*}(\phi_{D}(\tilde{\mathbf{p}}))\mathbf{w}_{j}$
is the dominant-path beam response, and $n_{j}\sim\mathcal{CN}(0,\sigma^{2})$
is the beam-wise independent measurement noise. The residual non-dominant
multipath components are treated as bounded model mismatch. The RSS
observation is thus given by
\begin{equation}
g_{j}(\tilde{\mathbf{p}})=|\beta_{D}|^{2}b_{j}(\phi_{D})+\varepsilon_{j}\label{eq:Var-2-1}
\end{equation}
where $\varepsilon_{j}\triangleq2\mathfrak{R}\{\mu^{*}_{j}n_{j}\}+|n_{j}|^{2}$
denotes the RSS perturbation in the energy domain.

First, we compute the mean $\mathbb{E}[\varepsilon_{j}]$ and the
variance $\mathrm{Var}(\varepsilon_{j})$. Since $n_{j}$ is zero-mean
complex Gaussian, $\mathbb{E}[\mathfrak{R}\{\mu^{*}_{j}n_{j}\}]=0$,
and $\mathbb{E}[|n_{j}|^{2}]=\sigma^{2}$, we can therefore have 
\begin{equation}
\mathbb{E}[\varepsilon_{j}]=\sigma^{2}\label{eq:Var-2}
\end{equation}
which shows that RSS measurements contain a noise bias $\sigma^{2}$.
To compute the variance, write $n_{j}=n_{R}+in_{I}$ with $n_{R},n_{I}\sim\mathcal{N}(0,\sigma^{2}/2)$.
Then $2\mathfrak{R}\{\mu^{*}_{j}n_{j}\}\sim\mathcal{N}(0,2\sigma^{2}|\mu_{j}|^{2})$.
Meanwhile, $|n_{j}|^{2}=n^{2}_{R}+n^{2}_{I}$ satisfies $\mathbb{E}[|n|^{2}_{j}]=\sigma^{2}$
and $\mathrm{Var}[|n|^{2}_{j}]=\sigma^{4}$. The terms $2\mathfrak{R}\{\mu^{*}_{j}n_{j}\}$
and $|n_{j}|^{2}$ are uncorrelated. Thus 
\begin{equation}
\begin{aligned}\mathrm{Var}(\varepsilon_{j}) & \approx2\sigma^{2}|\mu_{j}|^{2}+\sigma^{4}=2\sigma^{2}|\beta_{D}|^{2}b_{j}(\phi_{D})+\sigma^{4}.\end{aligned}
\label{eq:Var}
\end{equation}
At high SNR, the term $\sigma^{4}$ is negligible, yielding $\mathrm{Var}(\varepsilon_{j})\approx2\sigma^{2}|\beta_{D}|^{2}b_{j}(\phi_{D})$.
Also, for different beams $j\neq j^{'}$, the noises are independent,
so $\varepsilon_{j}$ are independent.

Second, we obtain an approximation of normalized RSS vector $\widetilde{\mathbf{g}}$
around the noiseless term $\mathbf{f}(\phi_{D})$. Rewrite $\mathbf{g}=|\beta_{D}|^{2}\mathbf{b}(\phi_{D})+\boldsymbol{\varepsilon}$
with $\boldsymbol{\varepsilon}=[\varepsilon_{j}]$. Define the scalar
\begin{equation}
S\triangleq\mathbf{1}^{T}\mathbf{g}=|\beta_{D}|^{2}B+E,\hspace{0.2cm}B\triangleq\mathbf{1}^{T}\mathbf{b}(\phi_{D}),\hspace{0.2cm}E=\mathbf{1}^{T}\boldsymbol{\varepsilon}.\label{eq:Var-1}
\end{equation}
Applying first-order Taylor expansion (Delta method) around the noiseless
term yields
\begin{equation}
\widetilde{\mathbf{g}}\approx\mathbf{f}(\phi_{D})+\boldsymbol{\eta},\hspace{0.2cm}\boldsymbol{\eta}\triangleq\frac{1}{|\beta_{D}|^{2}B}(\boldsymbol{\varepsilon}-\mathbf{f}(\phi_{D})\mathbf{1}^{T}\boldsymbol{\varepsilon}).\label{eq:Var-1-1}
\end{equation}
Since $\mathbb{E}[\boldsymbol{\varepsilon}]=\sigma^{2}\mathbf{1}$,
we can have 
\begin{equation}
\mathbb{E}[\boldsymbol{\eta}]=\frac{\sigma^{2}}{|\beta_{D}|^{2}B}(\mathbf{1}-\mathbf{f}(\phi_{D})\mathbf{1}^{T}\mathbf{1}).\label{eq:Var-1-1-1}
\end{equation}
The normalized RSS is not exactly unbiased at finite SNR due to the
noise floor $\sigma^{2}$, but the bias magnitude is $||\mathbb{E}[\boldsymbol{\eta}]||=\mathcal{O}(1/\mathrm{SNR})$,
which is asymptotically unbiased as $\mathrm{SNR}\shortrightarrow\infty$.
Define $\boldsymbol{\Sigma}_{\varepsilon}=\mathrm{Cov}(\boldsymbol{\varepsilon})$
and $\mathbf{A}\triangleq\mathbf{I}-\mathbf{f}(\phi_{D})\mathbf{1}^{T}$.
Then $\boldsymbol{\eta}=1/(|\beta_{D}|^{2}B)\mathbf{A}\boldsymbol{\varepsilon}$.
Hence the covariance becomes 
\begin{equation}
\boldsymbol{\Sigma}_{\eta}\triangleq\mathrm{Cov}(\boldsymbol{\eta})=\frac{1}{|\beta_{D}|^{4}B^{2}}\mathbf{A}\boldsymbol{\Sigma}_{\varepsilon}\mathbf{A}^{T}.\label{eq:Var-1-1-1-1}
\end{equation}
At high SNR, using $\boldsymbol{\Sigma}_{\varepsilon}\approx2\sigma^{2}|\beta_{D}|^{2}\mathrm{diag}(\mathbf{b}(\phi_{D}))$,
then
\begin{equation}
\boldsymbol{\Sigma}_{\eta}\approx\frac{2\sigma^{2}}{|\beta_{D}|^{2}}\cdotp\frac{1}{B^{2}}\mathbf{A}\mathrm{diag}(\mathbf{b}(\phi_{D}))\mathbf{A}^{T}.\label{eq:Var-1-1-1-1-1}
\end{equation}
So the normalized RSS perturbation scales like $\mathcal{O}(1/\mathrm{SNR})$.

The MLE of $\phi_{D}$ is defined as
\begin{equation}
\hat{\phi_{D}}=\underset{\phi_{D}}{\mathrm{argmin}}(\widetilde{\mathbf{g}}-\mathbf{f}(\phi_{D}))^{T}\boldsymbol{\Sigma}^{-1}_{\eta}(\widetilde{\mathbf{g}}-\mathbf{f}(\phi_{D})).\label{eq:Var-1-1-1-1-1-1}
\end{equation}
Assume that $\mathbf{f}(\phi_{D})$ is differentiable in a neighborhood
of the true AOD and satisfies the local identifiability condition
\begin{equation}
\mathbf{f}'(\phi_{D})^{T}\boldsymbol{\Sigma}^{-1}_{\eta}\mathbf{f}'(\phi_{D})>0.\label{eq: linear approximation-1}
\end{equation}
This condition ensures that a small change in AOD induces a distinguishable
change in the normalized beam-power pattern.

Using first-order Taylor expansion, we can have 
\begin{equation}
\hat{\phi}_{D}-\phi_{D}\approx\frac{\mathbf{f}'(\phi_{D})^{T}\boldsymbol{\Sigma}^{-1}_{\eta}\boldsymbol{\eta}}{\mathbf{f}'(\phi_{D})^{T}\boldsymbol{\Sigma}^{-1}_{\eta}\mathbf{f}'(\phi_{D})}.\label{eq: linear approximation}
\end{equation}
Taking expectation gives
\begin{equation}
\mathbb{E}[\hat{\phi}_{D}-\phi_{D}]\approx\frac{\mathbf{f}'(\phi_{D})^{T}\boldsymbol{\Sigma}^{-1}_{\eta}\mathbb{E}[\boldsymbol{\eta}]}{\mathbf{f}'(\phi_{D})^{T}\boldsymbol{\Sigma}^{-1}_{\eta}\mathbf{f}'(\phi_{D})}=\mathcal{O}(1/\mathrm{SNR}).\label{eq:Man_linear approximation}
\end{equation}
Thus, as $\mathrm{SNR}\rightarrow\infty$, $\mathbb{E}[\hat{\phi}_{D}-\phi_{D}]\shortrightarrow0$,
i.e., asymptotically unbiased. Furthermore, 
\begin{equation}
\mathrm{Var}(\hat{\phi}_{D}-\phi_{D})\approx\frac{1}{\mathbf{f}'(\phi_{D})^{T}\boldsymbol{\Sigma}^{-1}_{\eta}\mathbf{f}'(\phi_{D})}.\label{eq:Var_linear approximation-1}
\end{equation}
Because $\boldsymbol{\eta}$ is approximately Gaussian, $\hat{\phi}_{D}-\phi_{D}$
is also Gaussian. Therefore,
\begin{equation}
\hat{\phi}_{D}\sim\mathcal{N}(\phi_{D},(\mathbf{f}'(\phi_{D})^{T}\Sigma^{-1}_{\eta}\mathbf{f}'(\phi_{D}))^{-1})\label{eq:AOD_model-1-1}
\end{equation}
which completes the proof.

\end{appendices}

\bibliographystyle{IEEEtran}
\bibliography{Bib/IEEEabrv,Bib/Reference_MIMO,Bib/Reference_Charting,Bib/Reference}

\end{document}